\documentclass[12pt]{extarticle}
\usepackage{james}

\customtitle{Investigating methods to solve large windfarm optimization
problems with a minimum number of qubits using
circuit-based quantum computers}{
James Hancock$^{1,*}$, Matthew J. Craven$^1$ and Craig McNeile$^1$.\\
$^1$School of Engineering, Computing and Mathematics, University of Plymouth\\ Drake Circus, Plymouth PL4 8AA, UK\\
$^*$Author to whom any correspondence should be addressed\\(\url{james.hancock@plymouth.ac.uk})}{\today}

\begin{document}
\maketitle

\begin{abstract}
    This study investigates quantum computing approaches for solving the
    windfarm layout optimization (WFLO) problems formulated as a quadratic
    unconstrained binary optimization (QUBO) problem. We investigate two
    encoding methods that require fewer than one qubit per grid point:
    the previously developed Pauli correlation encoding (PCE) and a novel
    single-qubit operator encoding (SQOE). These methods are tested on
    three windfarm configurations - two from prior WFLO scaling studies
    and a new real-world model based on an existing windfarm in Wales.
    The improved encoding methods allow us to solve WFLO problems on $9\times 9$ grids using up to 20 qubits on a quantum computer simulator. The results show that both encoding methods perform competitively and
    demonstrate favorable scaling characteristics across the tested
    systems.
\end{abstract}

\section{Introduction}
\subsection{General introduction}
As global energy demand rises, maximizing extraction from renewable sources is crucial to mitigate environmental impact. The UK's Climate Change Act \cite{climatechangeact2008} mandates net-zero emissions by 2050, calling for the scaling of the renewable energy infrastructure. Wind power, which already contributes $29.4\%$ of the UK's total electricity generation \cite{nationalgrid2023renewable}, is a key component of this transition due to its scalability and relatively low costs. Turbine placement is a critical optimization problem for maximizing windfarm power output, because a turbine that is too close to another will have reduced windspeed and thus generate less power.
The optimization problem to find the optimal configuration of wind turbines in a windfarm
is the windfarm layout optimization (WFLO) problem.

While the primary focus of this study is to optimize windfarm layouts with a power output-based cost function, it is also important to consider the associated economic implications. Inefficient windfarm layouts translate directly into financial losses for developers and operators. It has been estimated that a power prediction error of only one GWh can result in annual revenue losses of €50,000–€70,000~\cite{wes-6-311-2021}. Over a 20-year project lifespan, this exceeds €2.4 million. Long-distance wake effects from neighboring or upstream windfarms, often called wind theft, can further reduce energy yield~\cite{bbc_wind_theft}, introducing discrepancies between actual and expected performance. These inter-farm wake interactions are not considered in this study, which focuses on internal wake dynamics within a single windfarm.

Classical approaches to windfarm layout optimization face scalability limitations due to the exponential size of the solution space. This motivates developing algorithms to solve the WFLO problem using quantum computers \cite{mine}. As reviewed in reference \cite{Abbas:2023agz}, the comparison of the performance of classical and quantum optimizers can be more subtle than just polynomial versus exponential time performance. There have been only a few studies on solving the WFLO problem using quantum computers
\cite{senderovich2022exploiting,mine,kagemoto2024possible,nigro2025leveraging}

In our initial study \cite{mine} of solving the WFLO problem using a circuit-based quantum computer, the windfarm was described as a discrete grid and each grid point controlled by one qubit. This required a large number of qubits to simulate large grid sizes. In this paper, we investigate mapping the grid points to correlated Pauli string operator expectation values following the method developed in reference \cite{Sciorilli2025}. We also develop and test a new method to map binary variables to Pauli operator expectation values. This mapping enables multi-parameter gradient estimation from a single set of counts using a configurable number of qubits, a method we label as single-qubit operator encoding (SQOE).

Many classical solvers for optimization problems use techniques to find starting values for the initial parameter values, which approximate a high-quality solution, known as warm starts \cite{john2008implementation}. The development of quantum algorithms for warm starts is currently an active area of investigation \cite{truger2024warm}. We investigate the use of warm starts to improve the convergence of our quantum solvers.

We tested the developed quantum solvers on two standard WFLO benchmark problems to study the scaling of the algorithms. As a case study, we also applied our quantum solvers to the Alltwalis windfarm in Wales \cite{thewindpower_alltwalis_2025}. This helps us investigate practical feasibility across different scales. We investigated grids ranging from $4\times4$ (16 binary variables) to $9\times9$ (81 binary variables). To assess our quantum solvers, we use a simplified model to analyze runtime scaling as the system size increases.

\subsection{Windfarm layout optimization}

Windfarm layout optimization (WFLO) determines optimal turbine configurations subject to multiple constraints, including wind regime, topography, turbine count, and economic, engineering, and environmental factors.

A turbine located downstream from another, within its cone of effect, experiences reduced wind speed and thus lower power output compared to a turbine in undisturbed flow. This effect is called a turbine's wake.  The key issue in turbine placement in the wind farm is minimizing energy loss due to wakes from other turbines.

Classically, the problem of WFLO is solved in several different ways, and using different representations. Possible representations and methods include using continuous, discrete, and binary variables. Continuous models (e.g., references \cite{GUIRGUIS2017279} and \cite{GUIRGUIS2016110}) allow for precise placement of turbines on the windfarm. Continuous models offer high accuracy at the cost of high computational expense. The cited studies employ both gradient-based optimization techniques and nonlinear programming approaches to address the WFLO problem. 


Through the use of discrete variables for WFLO, several algorithmic approaches have been developed. Manikowski et al. \cite{jmse9121376} introduced a multi-objective model using a hill-climbing algorithm to balance economic considerations (turbine cost) and power output of fixed offshore windfarms, later expanding their analysis to include objectives such as windfarm efficiency, annual energy production, and levelized cost of energy \cite{MANIKOWSKI2025112879}; the study also included continuous variables. Further discrete methods include a complex-valued code pathfinder algorithm \cite{LI2022100307}, a learning-enhanced genetic algorithm \cite{dong2024reinforcementlearningenhancedgeneticalgorithm}, and multi-objective variable reduction techniques \cite{RAMLI2023101016}. Algorithms based on the Quota Steiner tree problem have demonstrated significant speed-ups compared to commercial solvers like Gurobi \cite{pedersen2025integrated}. Within this landscape of discrete optimization, our work focuses specifically on the quadratic unconstrained binary optimization (QUBO) formulation of the WFLO problem.

Senderovich et al. \cite{senderovich2022exploiting} showed that the formulation of the WFLO we investigate is a nondeterministic polynomial time (NP)-hard problem. We do not expect that a quantum solver will solve the WFLO problem in polynomial time. To establish the potential advantage of quantum algorithms over classical solvers, care must be taken in the study.
\section{The WFLO model}\label{sec:WFLO}
\subsection{Quadratic unconstrained binary optimization}\label{sec:WFLOQUBO}
QUBO is a set of nondeterministic NP-hard problems \cite{QUBOsurvey}, onto which many combinatorial optimization problems can be mapped. A QUBO problem is defined as:
\begin{equation}
    \underset{x}{\text{argmin }}f_Q(x)\text{,}
\end{equation}
where
\begin{equation}\label{eq:xQx}
    f_Q(x) := x^TQx = \sum_{i,j = 0}^NQ_{ij}x^ix^j,
\end{equation}
and $Q$ is a real symmetric or upper-triangular matrix 
(called the QUBO matrix) and $x\in\{0,1\}^N$ is a binary vector.  

\subsection{Mapping the WFLO to a QUBO problem}
We further develop the mapping of the WFLO to a QUBO model, first proposed by Senderovich et al. in reference \cite{senderovich2022exploiting}, by introducing an additional term that allows for undesirable sites on the grid. 

The key WFLO challenge is mitigating wake effects that reduce windspeeds. We address this by optimizing layouts using annual wind regimes. A wind regime is a probability distribution over all possible windspeed and direction combinations.

\begin{figure}[H]
  \centering
  \begin{subfigure}[t]{0.48\textwidth}
    \centering
    \includegraphics[width=\linewidth]{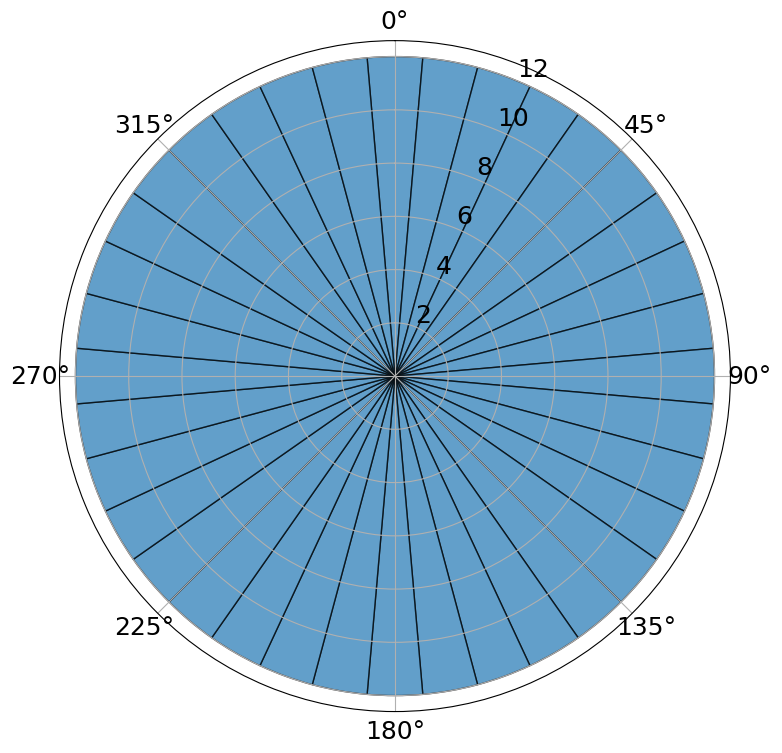}
    \caption{Rose diagram of $D_{SWR}$ windspeeds.}
    \label{fig:myRoseSpeed}
  \end{subfigure}
  \hfill
  \begin{subfigure}[t]{0.48\textwidth}
    \centering
    \includegraphics[width=\linewidth]{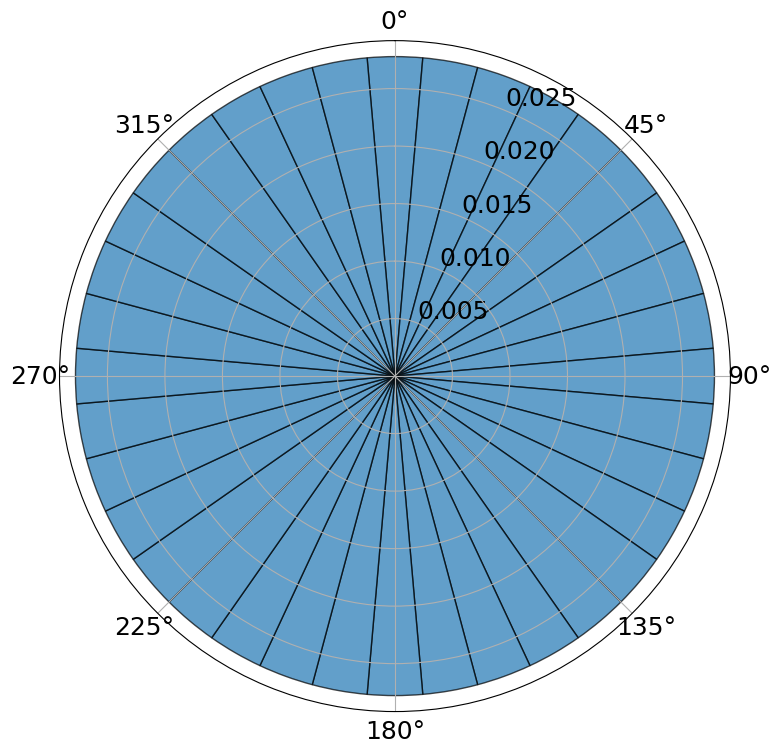}
    \caption{Rose diagram of $D_{SWR}$ data probabilities.}
    \label{fig:myRoseProb}
  \end{subfigure}
  \caption{Rose diagrams showing windspeed and probability distributions for the second Mosetti benchmark wind regime.}
\end{figure}

A common benchmark for WFLO problems is the Mosetti wind regimes \cite{MOSETTI1994105}, which we use here as an illustrative example. The second benchmark regime is defined by a windspeed of $12\text{ms}^{-1}$ spaced every $10$ degrees, all with equal probability. This regime can be represented in multiple ways: by rose diagrams for speed and probability (\fref{fig:myRoseSpeed} and \fref{fig:myRoseProb}, respectively), in a table (\tref{tab:dwr_windregime}), or as a set of arrangements:
\begin{equation}
D_{SWR} = \left\{\left\{10k,12\text{ms}^{-1},\frac{1}{36}\right\}\right\}_{k=0}^{35} = \left\{\left\{\alpha_d,v_d,p_d\right\}\right\}_{d\in D}.
\end{equation}
In this definition, the inner sets are the wind arrangements ($d$) and the full set is the wind regime ($D$). Each arrangement $d$ comprises the wind direction $\alpha_d$, free windspeed $v_d$, and probability of occurrence $p_d$.

\begin{table}[H]
  \centering
  \begin{tabular}{ccc}
    \toprule
    $\alpha_d$ & $v_d$ & $p_d$ \\
    \midrule
    $0$   & 12 & 0.0278 \\
    $10$  & 12 & 0.0278 \\
    $\vdots$ & $\vdots$ & $\vdots$ \\
    $340$ & 12 & 0.0278 \\
    $350$ & 12 & 0.0278 \\
    \bottomrule
  \end{tabular}
  \caption{$D_{SWR}$, the second Mosetti benchmark wind regime. $\alpha_d$ is the angle, $v_d$ is the free windspeed and $p_d$ is the probability of a given arrangement $d$.}
  \label{tab:dwr_windregime}
\end{table}

To model wake interactions according to the wind regime, we use the linear superposition of wakes (LS) method \cite{Donovan2005WindFO}. While other models, such as sum-of-squares \cite{SSproceedings} or non-linear formulations (appendix \ref{appen:KJWM}), are often used, the QUBO setting requires a linear construction to leverage quantum hardware. We therefore opt for the LS method. 

For an $ L\times L$ grid with $N$ total sites, the power output for turbines placed at sites $J\subseteq\{1,\dots, N\}$, calculated via LS, is given by
\begin{equation}\label{eq:ELS}
    W^{\text{LS}} = \sum_{d\in D}\sum_{i\in J} p_d\left[\frac{1}{3}v_d^3 - \sum_{j\in w_i}\frac{1}{3}\left(v_d^3 - u_{ij}^3\right)\right],
\end{equation}
where $w_i$ is the set of sites which are in the wake of a turbine at $i$ and $u_{ij}$ is the reduced windspeed at $j$ due to the turbine at $i$.

The radius of a turbine's effective wake cone depends on several physical factors. We can calculate the wake radius at a distance $\delta$ as
\begin{equation}
    r_w = r_t + a\delta,
\end{equation}
where $r_t$ is the turbine radius and $a$ is the wake expansion factor. 

To calculate reduced windspeed, we use the simplified Jensen top-hat deficit \cite{3a81166868144671af170595fd17b8f6} 
\begin{equation}
    \frac{u_{ij}}{v_d} = \left(1-\sqrt{1-C_T}\right)\left(\frac{r_t}{r_t+a\delta}\right)^2=\left(1-\sqrt{1-C_T}\right)\left(\frac{r_t}{r_w}\right)^2,
\end{equation}
where $C_T$ is the turbine thrust coefficient.

We can reformulate this model as a QUBO problem by introducing binary variables $x^i \in \{0,~1\}$, where $x^i = 0$ indicates the absence of a turbine and $x^i = 1$ represents its presence. The mapping yields the following cost function:
\begin{equation}\label{eq:Energy}
    f(x) = (-1)\sum_{d\in D}\sum_{i=1}^{N} p_d\left[\frac{1}{3}v_d^3x^ix^i - \sum_{j\in w_i}\frac{1}{3}\left(v_d^3 - u_{ij}^3\right)x^ix^j\right],
\end{equation}
where the $(-1)$ factor comes from our minimization formulation.

The current cost function $f(x)$ leads to an optimal solution where a turbine is placed at every site, which is unrealistic. To improve the model, we introduce three additional constraints: a maximum number of $M\in\{0,1,..., N\}$ turbines, a minimum spacing of $E\in \left[0,\sqrt2L\right]$ (where $L$ is the length of the longest side of the grid) between turbines, and unfavorable sites, which we store as a vector $\vec{P}$.

We limit the number of turbines using the term
\begin{equation}
    c_1(x,M) = \left(\sum_{i=1}^{N}x^i - M\right)^2.
\end{equation}
The proximity constraint is included through the term
\begin{equation}
    c_2(x,E) = \sum_{\lVert x_\text{phys}^i-x_\text{phys}^j\rVert_2<E}x^ix^j,
\end{equation}
where $x_\text{phys}^i$ is the Cartesian coordinate label of $x^i$, defined, along with $E$, in meters. Finally, the undesirable locations are defined by the term
\begin{equation}
    c_3(x,\vec{P}) = \sum_{i=1}^{N} P_ix^i = \vec{P}\cdot x,    
\end{equation}
which tells us which sites we do not want, with $P_i=0$ meaning acceptable, $P_i=1$ meaning maximal avoidance.

We add these constraints to $f(x)$, in the form of penalty terms, each weighted by a controllable parameter. Our full cost function then becomes
\begin{equation}
\begin{aligned}
    C(x,D,M,E,\vec{P};\lambda_1,\lambda_2,\lambda_3) &= f(x) + \lambda_1c_1(x,M) + \lambda_2c_2(x,E) + \lambda_3c_3(x,\vec{P}),
\end{aligned}
\label{eq:FinalCost}
\end{equation}
and our problem can be stated as finding
\begin{equation}
    x^*=\underset{x}{\text{argmin }}C(x,\dots).
\end{equation}

\subsection{Structure of the QUBO matrix}

The performance of algorithms to solve QUBO problems depends on the
structure of the QUBO matrix. Some special QUBO matrices can be solved in polynomial time \cite{Cela2022}. The structure of the QUBO matrix can be changed by modifying the implementation of the constraints \cite{liu2025framework}.

We analyze this in both functional form and the heatmap of the example QUBO matrices produced. The heatmap plots use a set of simple problem parameters (Windfarm $A$ \cite{rod2016} in \tref{tab:summaryText} in \sref{sec:Test_models}, with a unidirectional wind regime) on a $10\times10$ grid. This simple system has wind regime $D = \{\{0,12,1\}\}$, wake radius $r_w = 82 + 0.094\delta$ at distance $\delta$, and a windfarm area of $W_A=15.49\text{km}^2$. For further details on model construction, see appendix \ref{appen:TM}.

In \fref{fig:hm0} and \fref{fig:hm2500}, we observe that across the operational range of $\lambda$ values (see appendix \ref{appen:Lambda} for details on the choice of scale for $\lambda$), the matrix is characterized by large diagonal values, with off-diagonal terms that decay with distance. For this analysis, we set the maximum number of turbines to $M=16$ and omit other constraints, as they do not significantly alter the matrix structure. When $\lambda=0$ (i.e., no constraints are applied), the QUBO matrix is not fully dense; the constraint terms are responsible for filling the matrix.

\begin{figure}[H]
  \centering
  \begin{subfigure}[t]{0.48\textwidth}
    \includegraphics[width=\linewidth]{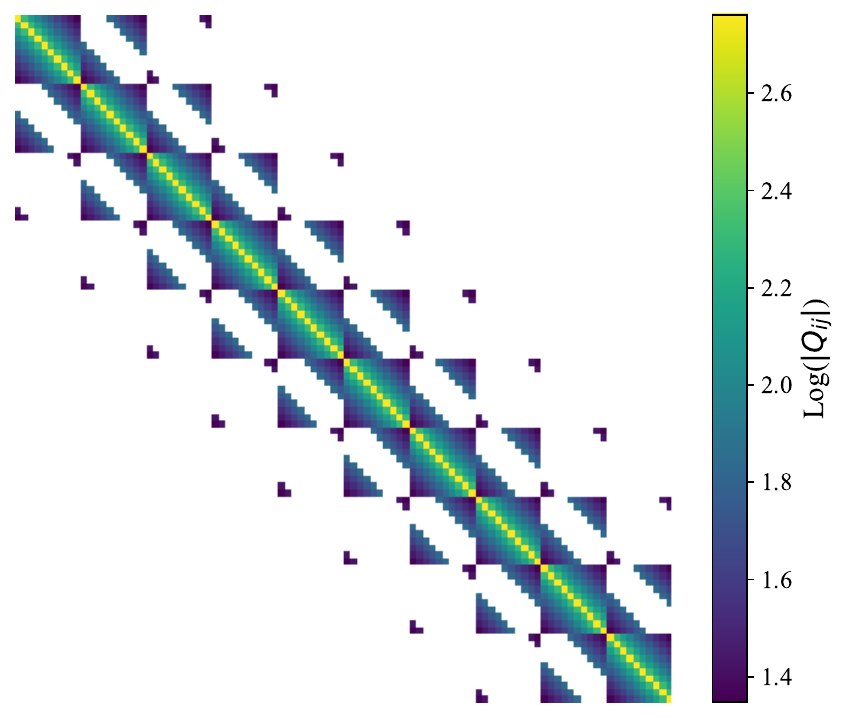}
    \caption{Heatmap of $\text{Log}$ of the absolute value of WFLO QUBO matrix with $\lambda=0$. White squares are entries $=0$.}
    \label{fig:hm0}
  \end{subfigure}
  \hfill
  \begin{subfigure}[t]{0.48\textwidth}
    \includegraphics[width=\linewidth]{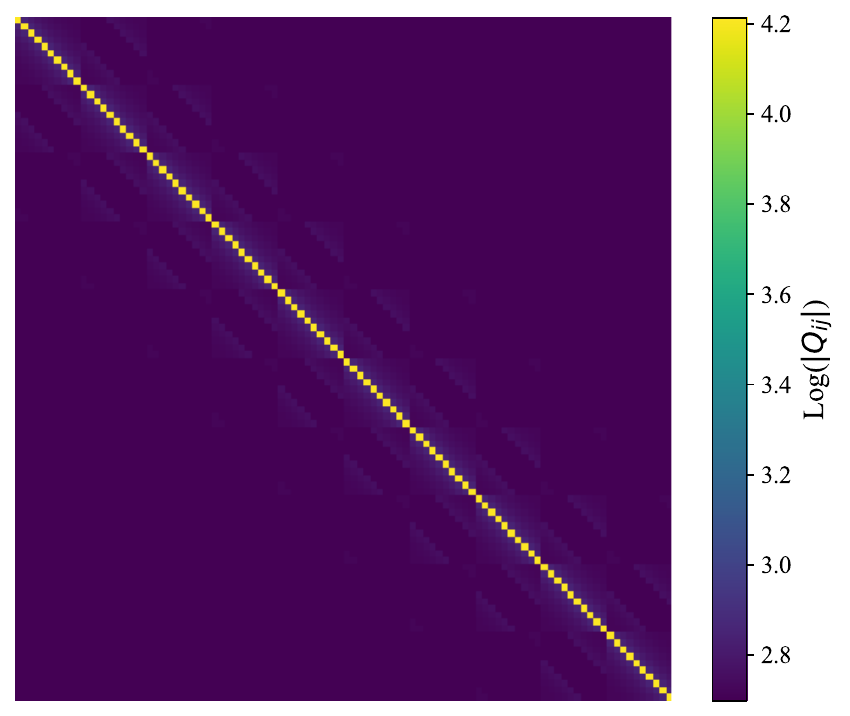}
    \caption{Heatmap of $\text{Log}$ of the absolute value of WFLO QUBO matrix with $\lambda=250$.}
    \label{fig:hm2500}
  \end{subfigure}
  \caption{Heatmaps of problem QUBO matrix with constraint weight (a) $\lambda=0$ and (b) $\lambda=250$.}
\end{figure}

Looking specifically at $f(x)$, we see it contains large terms along the diagonal, as well as sparse $O(L)$ connections, which account for wake effects. To understand how these terms generate $O(L)$-length connections, it is informative to examine a numbered grid, with the area of effect of a simplified wake from a single turbine, as shown in \fref{fig:gridlabelling_10x10_colored}. We can see that given a simple wind regime ($D=\{\{0,v,1\}\}$), a turbine at site $15$ leads to some sparse $O(L)$-length connections due to how we label the grid with binary vector variables. With a more complex wind regime, these connections act in all directions.

\begin{figure}[tbp]
  \centering
  \begin{tikzpicture}[scale=0.7]
    \draw[step=1cm, black, very thin] (0,0) grid (10,10);
    \foreach \y in {0,...,9} {
        \foreach \x in {0,...,9} {
            \pgfmathtruncatemacro{\labelnumber}{\x + 10*(9-\y) + 1}
            \draw (\x+0.5, \y+0.5) node{\labelnumber};
        }
    }
    \fill[red, opacity=0.5] (4,9) rectangle (5,8);
    \foreach \x/\y in {
        3/7, 4/7, 5/7, 
        2/6, 3/6, 4/6, 5/6, 6/6,
        1/5, 2/5, 3/5, 4/5, 5/5, 6/5, 7/5
    } {
        \fill[gray!50, opacity=0.5] (\x,\y+1) rectangle (\x+1,\y);
    }
  \end{tikzpicture}
  \caption{Labelling of sites on a $L=10$ windfarm grid. There is a turbine located at position 15. The wind regime is $D=\{\{0,v,1\}\}$, with capped wake length $3$ and wake radius $1$. These values are not realistic, but show the connections that occur due to wakes. More realistic wake patterns are shown in appendix \ref{appen:Wakes}.}
  \label{fig:gridlabelling_10x10_colored}
\end{figure}
\FloatBarrier

\begin{figure}[H]
  \centering
  \begin{subfigure}[t]{0.48\textwidth}
    \includegraphics[width=\linewidth]{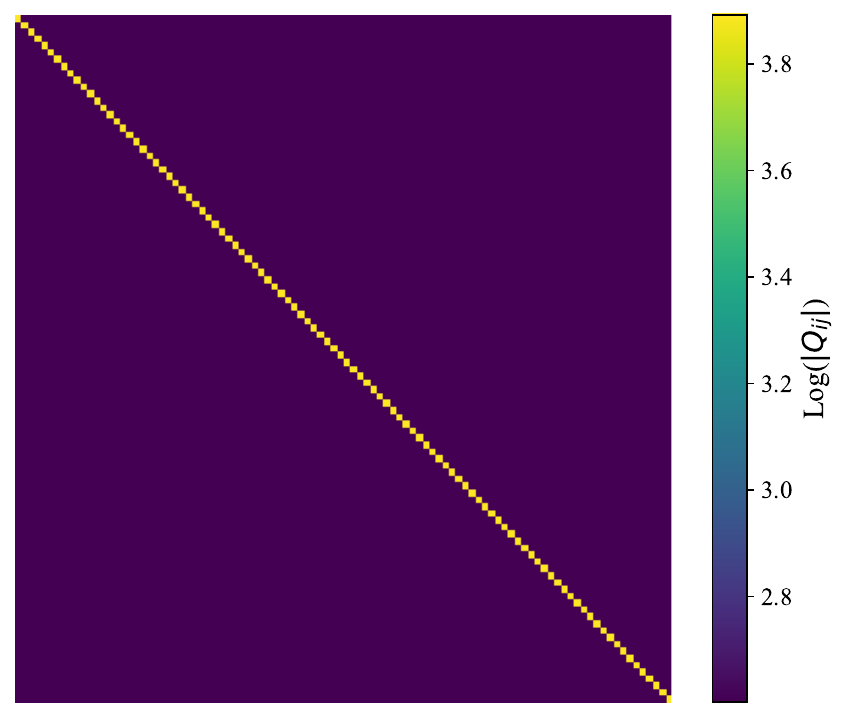}
    \caption{Heatmap of $\text{Log}$ of the absolute value of WFLO QUBO $M=4.0$ constraint matrix with $\lambda=200$.}
    \label{fig:hmM}
  \end{subfigure}
  \hfill
  \begin{subfigure}[t]{0.48\textwidth}
    \includegraphics[width=\linewidth]{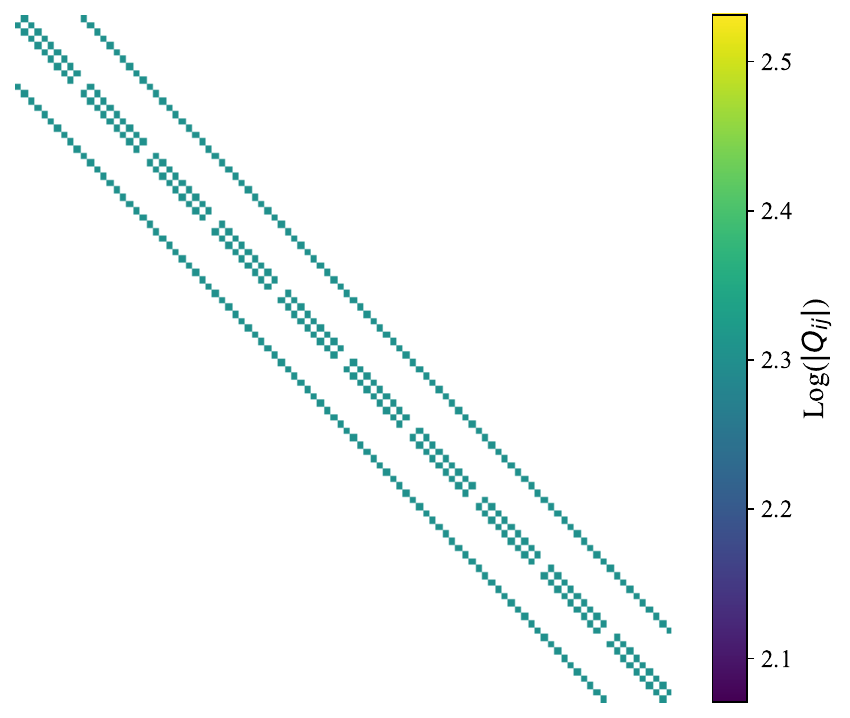}
    \caption{Heatmap of $\text{Log}$ of the absolute value of WFLO QUBO $E=465.0$\,m constraint matrix with $\lambda=200$.}
    \label{fig:hmE}
  \end{subfigure}
  \caption{Heatmaps of QUBO matrix for (a) $M=4.0$ constraint and (b) $E=465.0\text{m}$, both with $\lambda = 200$.}
\end{figure}

The $c_1$ term in equation \ref{eq:FinalCost}  is a dense, but simple-to-solve-for term. Solving this with a cold start (i.e., from a random initialization) is difficult; however, solvers can exploit the matrix structure (a constant value on the diagonal and another on all off-diagonal entries) to make a better initial guess. \Fref{fig:hmM} shows the heatmap of the constraint matrix for $M=10$.

\begin{figure}[tbp]
  \centering
  \begin{subfigure}[t]{0.48\textwidth}
    \begin{tikzpicture}[scale=0.7]
      \draw[step=1cm, black!30, very thin, opacity=0.7] (0,0) grid (10,10);
      \fill[red, opacity=0.5] (4,9) rectangle (5,8);
      \fill[red, opacity=0.5] (5,6) rectangle (6,5);
      \draw (4.5,8.5) node{\textbf{15}};
      \draw (5.5,5.5) node{\textbf{36}};
      \draw[<->, thick, red, line width=1.2pt]
        (4.5,8.5) -- (5.5,5.5)
        node[pos=0.6, below=12pt, sloped, black]
        {$\lVert x_\text{phys}^{15} - x_\text{phys}^{36} \rVert < E = 3$};
    \end{tikzpicture}
    \caption{Turbines at sites 15 and 46 on an $L=10$ grid, connected when $E=3$ (boxes). 
             The red arrow indicates their interaction range.}
    \label{fig:gridlabelling_final}
  \end{subfigure}
  \hfill
  \begin{subfigure}[t]{0.48\textwidth}
    \includegraphics[width=\linewidth]{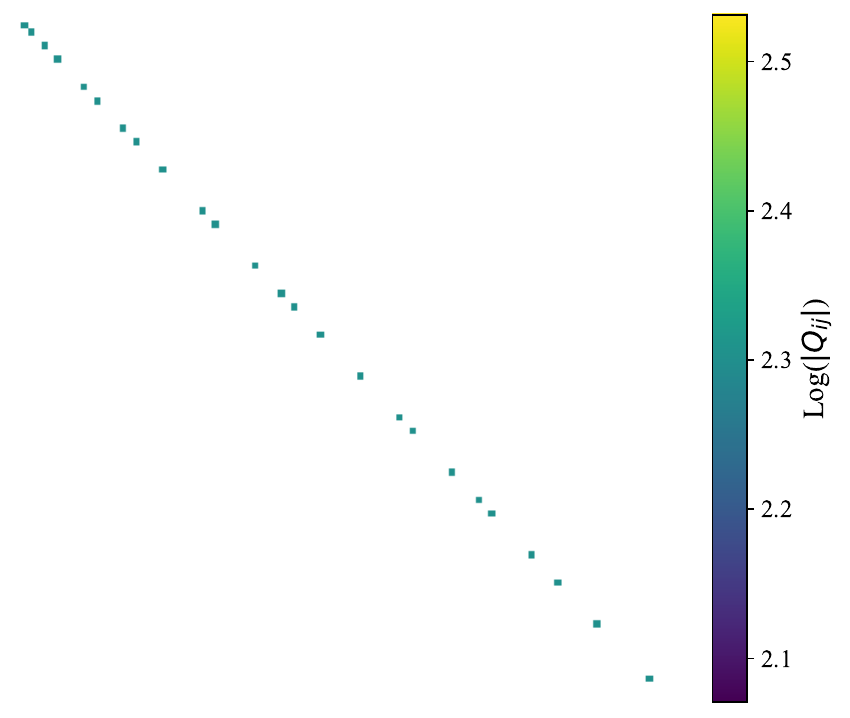}
    \caption{Heatmap of $\text{Log}$ of the absolute value of WFLO QUBO $\vec{P}$ constraint matrix, set such that $P_i=1$ if $i$ is a prime constraint matrix with $\lambda=200$.}
    \label{fig:hmP}
  \end{subfigure}
  \caption{(a) On grid diagram to show how minimum spacing constraint causes $O(L)$-length connections in QUBO matrix. (b) Heatmaps of QUBO matrix for (a) $\vec{P}$ constraint with $\lambda = 200$.}
\end{figure}
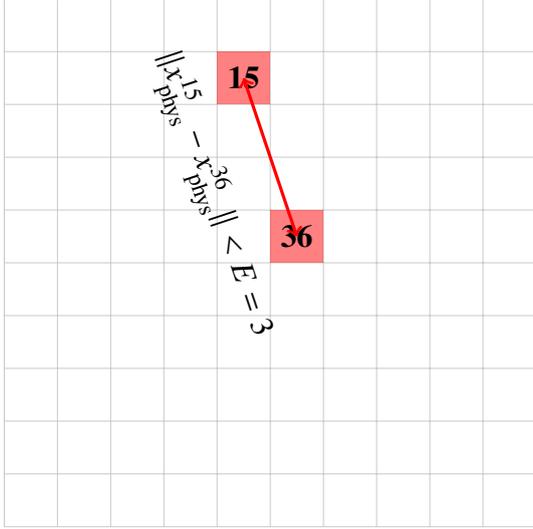
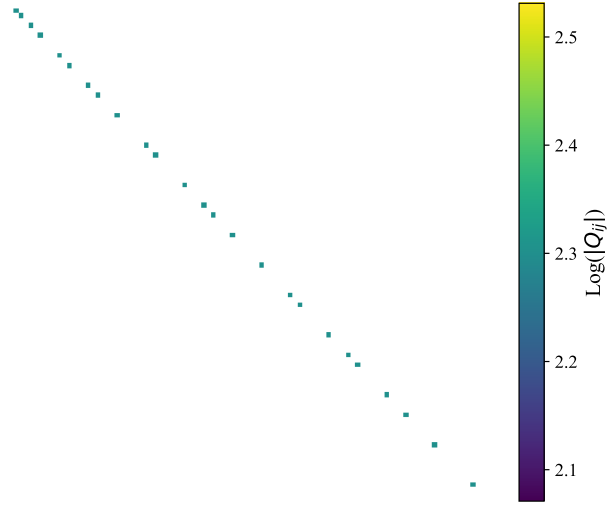

The $c_2$ term in equation \ref{eq:FinalCost} provides another sparse $O(L)$-length connection term. \Fref{fig:hmE} shows the heatmap of the constraint matrix for $E=465.0$m. In \fref{fig:gridlabelling_final}, we see how the minimum distance term leads to $O(L)$-length connections, similar to what we see in $f(x)$.

The $c_3$ term in equation \ref{eq:FinalCost} is a simple term, adding weights to certain values along the diagonal. \Fref{fig:hmP} shows the heatmap of the constraint matrix for $\vec{P}$ set such that $P_i=1$ if $i$ is a prime.

The QUBO matrix structure exhibits strong dependence on wake interactions and physical constraints specific to a given windfarm. This inherent variability problem-to-problem poses challenges for problem simplification approaches. For these reasons, we retain the full QUBO matrix in its current form without applying any simplifications.
\FloatBarrier

\subsection{Test Models}\label{sec:Test_models}
We assess quantum WFLO methods by analyzing both power output (solution quality) and computation time (efficiency). By evaluating these metrics across different problem scales, we assess the methods' scalability and practical usability for both current NISQ devices and future quantum computing capabilities.

To do this, we test two sets of models. The first is a simple toy model developed by Rodrigues et al. in reference \cite{rod2016}. The second is a novel construction of a real windfarm, modeled at different scales.

The key features of the models are summarized in \tref{tab:summaryText}. Full model construction details are provided in appendix \ref{appen:TM}.

\begin{table}[H]
    \centering
    \begin{tabular}{lrrrrr}
        \toprule
        Model & $r_t$ (m) & $h$ (m) & Wake equation & $W_A$ (km$^2$)\\
        \midrule
        Windfarm $A$ &  82 & 107 & $r_w=82 + 0.094\delta$ & 15.49\\
        Windfarm $B$ &  82 & 107 & $r_w=82 + 0.094\delta$ & 61.97\\
        Alltwalis & 46.5 & 90 & $r_w=46.5 + 0.154\delta$ & 1.5\\
        \bottomrule
    \end{tabular}
    
    \vspace{0.5cm}
    
    \begin{tabular}{lrrrrr}
        \toprule
        Model & Wind regime & $L$ & $M$ & $E$ (m) & $\vec{P}$ \\
        \midrule
        Windfarm $A$ & \Tref{tab:literature_windregime} & 4, 7, 9 & 16 & - & - \\
        Windfarm $B$ & \Tref{tab:literature_windregime} & 7, 9 & 49 & - & - \\
        Alltwalis & \Tref{tab:allt_windregime} & 7, 8, 9 & 10 & 465 & Appendix \ref{appen:P}\\
        \bottomrule
    \end{tabular}
    \caption{Parameters for our test models. The data for windfarms $A$ and $B$ originate from reference \cite{rod2016}. Alltwalis farm details can be found in reference \cite{thewindpower_alltwalis_2025}. Here, $r_t$ is the turbine radius, $h$ is the hub height, and $W_A$ is the area of the windfarm. \Tref{tab:literature_windregime} contains the North Sea wind data and \tref{tab:allt_windregime} the Alltwalis windfarm data.}
    \label{tab:summaryText}
\end{table}
\section{Quantum optimization methods}
In this study, we investigate two quantum circuit approaches for solving QUBO problems: Pauli correlation encoding (PCE) \cite{Sciorilli2025} and SQOE. Both methods encode spin variables through expectation values of operators. While PCE employs a black-box ansatz, SQOE uses an efficient single-parameter-per-qubit structure. This simplified ansatz enables access to heuristics, local gradient calculations, and, through multi-operator measurement, simultaneous gradient evaluation in a single set of counts.

\subsection{Quadratic unconstrained spin optimization}\label{sec:QUSO}
When solving these problems using quantum methods, directly controlling binary variables (i.e., values fixed at 0 or 1) can be difficult. Instead, it is often more convenient to recast the problem in terms of sign-based spin variables, where each binary variable is represented by a new variable $s^i \in \{-1,+1\}$. The mapping between binary and spin variables is given by
\begin{equation}\label{eq:mappingQUSO}
x^i \mapsto \frac{1}{2}\left(1 + s^i\right).
\end{equation}
Recalling the general QUBO formulation
\begin{equation}
f_Q(x) = \sum_{i,j}Q_{ij}x^ix^j,
\end{equation}
where $x^i \in {0,1}$, we can transform this into a quadratic unconstrained spin optimization (QUSO) form, using the substitution from equation \ref{eq:mappingQUSO}. QUSO problems are often referred to as Ising models, or problems defined by an Ising Hamiltonian. This yields:
\begin{equation}
\begin{aligned}
f_H(s) &= \sum_{i,j}\frac{1}{4}Q_{ij}\left(1+s^i\right)\left(1+s^j\right) \\
&= \sum_{i,j}\frac{1}{4}Q_{ij}\left(1 + s^i + s^j + s^is^j\right).
\end{aligned}
\end{equation}
Note that the constant term (proportional to 1) in the expansion does not affect the optimization and can therefore be omitted. Our QUSO cost is
\begin{equation}
\begin{aligned}
f_H(s) &= \sum_{i,j}\frac{1}{4}Q_{ij}\left(s^i+s^j+s^is^j\right)\\
&= \sum_{i,j}H_{ij}s^is^j + \sum_i h_is^i.
\end{aligned}
\label{eq:firstQMmodel}
\end{equation}
Equation~\ref{eq:firstQMmodel} was used in our initial study \cite{mine},
where one grid point is mapped to one qubit, and thus restricted the size of the grids that could be used.

In this study, we will use the expected value of quantum operators to hold our spin variable values. For each Pauli operator $P^i$, we define the spin variable as:
\begin{equation}
s^i = \mathrm{step}(t\langle P^i\rangle),
\end{equation}
where $t \in (0,\infty)$ is a tunable parameter that interpolates the $\text{step}$ function between a smoothed and a stepped sign function. We use $\tanh$ as our $\text{step}$ function.

\subsection{Pauli correlation encoding}
The first operator encoding methodologies we investigated were PCE \cite{Sciorilli2025,Carmo:2025tqt}.
In this, multi-qubit correlated operators are used to map many binary variables to few qubits. Given a system of $n$ qubits, we have a controllable parameter $k$ so that we can get $N=3{n\choose k}$ total spin variables.

We define the sets
\begin{equation}
\begin{aligned}
\tilde{X}_k^n &= \{X^{\otimes k}\otimes I^{\otimes(n-k)}, \text{ and its qubit-wise permutations}\},\\
\tilde{Y}_k^n &= \{Y^{\otimes k}\otimes I^{\otimes(n-k)}, \text{ and its qubit-wise permutations}\},\\
\tilde{Z}_k^n &= \{Z^{\otimes k}\otimes I^{\otimes(n-k)}, \text{ and its qubit-wise permutations}\},
\end{aligned}
\end{equation}
where $X,~Y,$ and $Z$ are the Pauli matrices and $I$ is the $2\times2$ identity matrix. These three sets contain all Pauli correlators with $k$ terms, many of which are highly non-local. As an example to help understanding, we can look at $\tilde{Z}_2^3$:
\begin{equation}
\begin{aligned}
\tilde{Z}_2^3 &= \{Z_0Z_1,~ Z_1Z_2,~Z_0Z_2\}\\
&= \{I\otimes Z\otimes Z,~ Z\otimes Z \otimes I,~Z\otimes I \otimes Z\},
\end{aligned}
\end{equation}
using Qiskit ordering.

Due to the non-local nature of these operators, we require a highly expressive, highly entangled ansatz. These must be treated as a black box. This method requires very few qubits (dependent on the choice of $k$), with the most efficient choice at each scale growing as $n\sim O(\log_2(N))$.

\subsection{Single-qubit operator encoding}
The second methodology that we explore is defined by considering the sets $\tilde{X}_1^n$ and $\tilde{Z}_1^n$, meaning $n=N/2$ continuous parameters can be used for $N$ binary variables.

The method itself was motivated by experiments using the PCE with IBM Quantum Computers provided by the UK National Quantum Computing Centre (NQCC) \cite{nqcc2024}. When using real hardware, queue times and circuit calls are often a bottleneck in the calculation of gradient (or simplex) values. The power in this mapping comes from the limited number of circuit parameters used - only one per qubit - which in turn limits the function calls required to calculate the gradient. To see why we need only one parameter, we can look at the matrix form of the Pauli operators for the $n=1$ case:
\begin{equation}
\begin{aligned}
X &= \begin{pmatrix}
0&1\\1&0
\end{pmatrix}\\
Z&=\begin{pmatrix}
1&0\\0&-1
\end{pmatrix},
\end{aligned}
\end{equation}
and a single qubit that has been acted on by a single $R_Y(\theta)$ rotation gate, which can be expressed in matrix-vector form as
\begin{equation}
\begin{pmatrix}
\cos\frac{\theta}{2} & -\sin\frac{\theta}{2}\\
\sin\frac{\theta}{2} & \cos\frac{\theta}{2}
\end{pmatrix}\begin{pmatrix}
1 \\ 0
\end{pmatrix} = \begin{pmatrix}
\cos\frac{\theta}{2}\\ \sin\frac{\theta}{2}
\end{pmatrix} = |\phi(\theta)\rangle.
\end{equation}
Taking the inner products with $X$ and $Z$, we see
\begin{equation}\label{eq:EVS}
\begin{aligned}
\langle \phi(\theta)|Z|\phi(\theta)\rangle = \cos^2\frac{\theta}{2}-\sin^2\frac{\theta}{2} &= \cos\theta =  Z(\theta)\\
\langle \phi(\theta)|X|\phi(\theta)\rangle = 2\cos\frac{\theta}{2}\sin\frac{\theta}{2} &= \sin\theta =  X(\theta).
\end{aligned}
\end{equation}

\begin{figure}[H]
\centering
\includegraphics[width=0.7\linewidth]{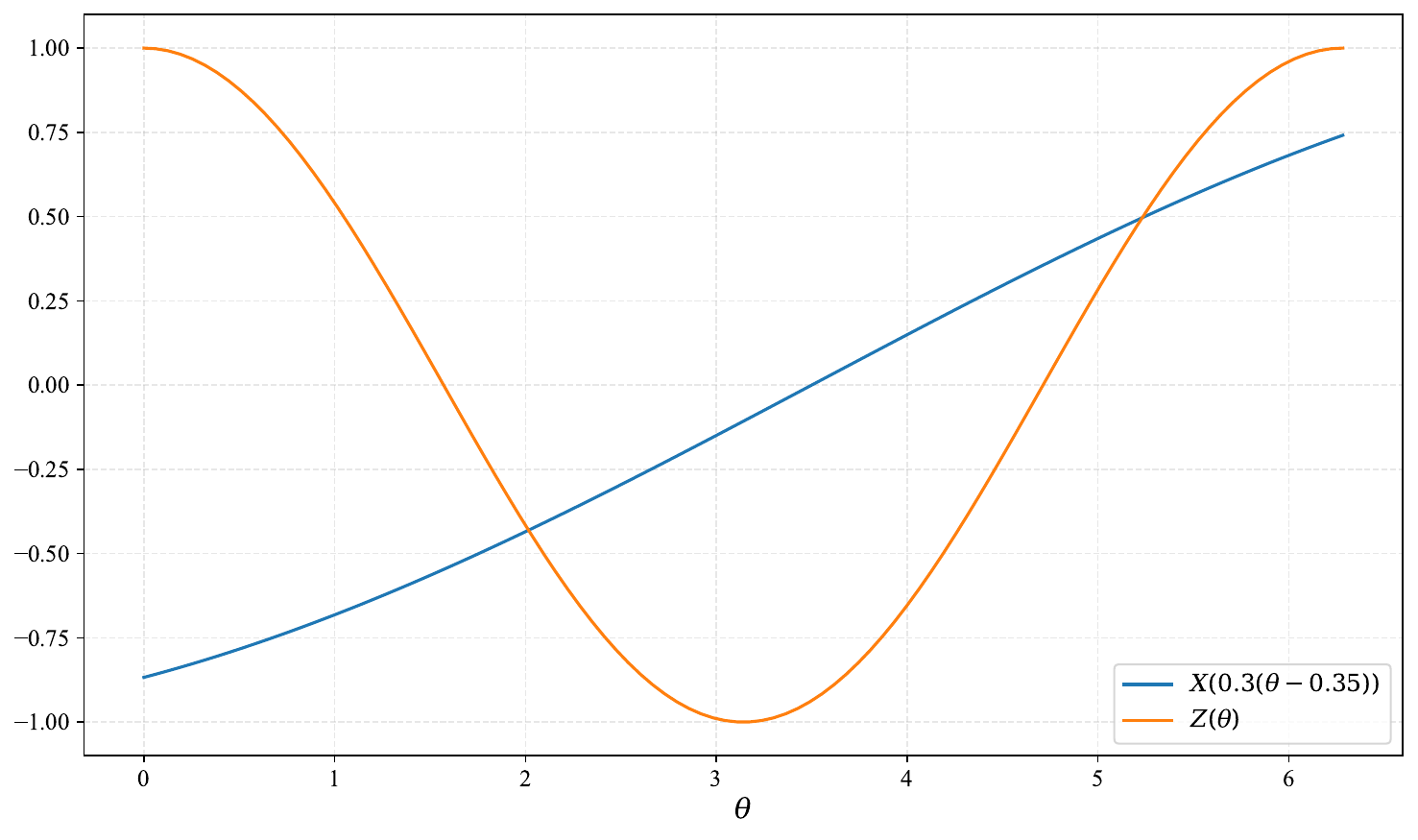}
\caption{Values of $Z(\theta)=\cos\theta$ (blue) and $X(0.3(\theta-3.5))=\sin\left(0.3(\theta-3.5)\right)$ (orange) over the range $[0,2\pi)$. We use these transformed versions to smooth the search space.}
\label{fig:transSig}
\end{figure}

We can map these to binary spin variables using the step function, as described in \sref{sec:QUSO},
\begin{equation}
\begin{aligned}
s^{j} &= \text{step}(t Z({\theta^i})),\\
s^{k} &= \text{step}(t X({\theta^i})),
\end{aligned}
\end{equation}
where the parameter $\theta^i$ controls two spin variables. The choice of the setup in equation \ref{eq:EVS} means that for varying choices of $\theta$, we can get all pairs of signatures $(s^{j},s^{k})\in{(\pm1,\pm1)}$. To simplify the optimization process, we instead use a transformed version of these operators, namely $Z(\theta)$ and $X(0.3(\theta-3.5))$, shown in \fref{fig:transSig}. This transformation aids the optimizer by simplifying the search space.

Using SQOE, we could use one qubit for all the variables, choosing a starting value, and then storing $\theta^i$ as it is updated. However, making use of multi-operator measurement methods (see \sref{sec:MOM}), we can measure $O(n)$ parameter gradients in the same evaluation over $n$ qubits. It is this ability to measure more than one function value simultaneously that offers an advantage.

\subsection{Local gradient calculation}\label{sec:LGC}
Consider a situation where we can change the variable $s^i$ continuously, independently of the rest. Drawing from statistical mechanics, we compute the gradient using only local variable changes. This reduces the computational cost from $O(N^2)$ to $O(N)$, as only the changing terms require evaluation. Changing the variable $s^i$ requires only the terms from $f_H$ that contain $\{H_{ij}\}_{j=0}^N$, $\{H_{ji}\}_{j=0}^N$, and $h_i$. In the symmetric representation, these are all terms in the row and column $i$ of $H$.

To visualize this, consider the example where we have $N=4$, and we can control $s^1$. Using the matrix form, we only need to consider terms that are in bold
\begin{equation}
\begin{pmatrix}
H_{00} & \mathbf{H_{01}} & H_{02} & H_{03}\\
\mathbf{H_{10}} & \mathbf{H_{11}} & \mathbf{H_{12}} & \mathbf{H_{13}}\\
H_{20} & \mathbf{H_{21}} & H_{22} & H_{23}\\
H_{30} & \mathbf{H_{31}} & H_{32} & H_{33}
\end{pmatrix}.
\end{equation}

With the SQOE, a change to one parameter modifies two variables; this means both their corresponding rows and columns must be considered. If the parameter $\theta^0$ controlled (for example) $s^0$ and $s^1$, we would consider the following terms
\begin{equation}
\begin{pmatrix}
\mathbf{H_{00}} & \mathbf{H_{01}} & \mathbf{H_{02}} & \mathbf{H_{03}}\\
\mathbf{H_{10}} & \mathbf{H_{11}} & \mathbf{H_{12}} & \mathbf{H_{13}}\\
\mathbf{H_{20}} & \mathbf{H_{21}} & H_{22} & H_{23}\\
\mathbf{H_{30}} & \mathbf{H_{31}} & H_{32} & H_{33}
\end{pmatrix}.
\end{equation}

\subsection{Multi-operator measurement}\label{sec:MOM}
When using quantum hardware, each set of calls to the quantum circuit requires a load time. This load time can be longer than some post-processing of the measurement data. In the aim of reducing load times, we investigated measuring multiple operators with one set of shots.

Consider the operator $P$, which may be a single or multi-qubit operator. The measurement protocol first applies the necessary unitary transformations to the circuit, then measures the target qubit(s) in the $ Z$-basis. This is repeated many times (for a number of shots) until we have a set of counts. The set of counts can be represented as $\{C_{B(i)}\}_{i=0}^{2^n-1}$, where $B(i)$ is the binary representation of $i$, e.g., $B(3) = 11$. Here $C_{B(i)}$ is the number of times we measured in the state $|B(i)\rangle$. $\sum_iC_{B(i)}=S$ is the total number of shots. Usually, we calculate the expected value of $P$ as
\begin{equation}
\langle P\rangle = \frac{1}{S}\sum_i(-1)^{w(B(i))}C_{B(i)},
\end{equation}
where $w(x)$ is the Hamming weight.

Now consider a second operator, $Q$, which can again be single or multi-qubit, but does not share any qubits with $P$. To calculate the expected value of both $P$ and $Q$ by measuring both operators simultaneously via creating slightly modified Hamming weight functions, $w_P(x)$ and $w_Q(x)$. These consider the Hamming weight across the qubits that are involved in $P$ and $Q$, respectively. Explicitly, the expected values can be calculated as:
\begin{equation}
\begin{aligned}
\langle P\rangle &= \frac{1}{S}\sum_i(-1)^{w_P(B(i))}C_{B(i)}\\
\langle Q\rangle &= \frac{1}{S}\sum_i(-1)^{w_Q(B(i))}C_{B(i)}.
\end{aligned}
\end{equation}

This can be generalized further to measure any number of operators, given they do not share any qubits. When real hardware is used, we do not want to measure operators on adjacent qubits as this can lead to crosstalk \cite{PRXQuantum.3.020301}.

By combining multi-operator measurements with the local gradient computation method from \sref{sec:LGC}, we can efficiently estimate gradients for multiple parameters simultaneously using a single set of quantum measurements, while avoiding the need to calculate the full cost.

\subsection{Classical Optimization Routine}
We optimize the PCE using the COBYLA algorithm \cite{COBYLA}, which leverages linear surrogate models for black-box optimization and has an efficient Python implementation.

When using the SQOE, we will use stochastic gradient descent (SGD) \cite{10.1214/aoms/1177729586}. This is similar to standard gradient descent, but the parameters whose gradient is calculated for each iteration are randomly selected. SGD is more efficient than standard (batch) gradient descent \cite{LeCun1998Efficient}. In addition, to help move around the complex search space, we employ a random value for $h$ in our central difference local gradient calculation:
\begin{equation}
\frac{\partial f_{H}(\theta)}{\partial \theta^i} = \frac{f_{H}(\theta + h\hat{\theta^i})-f_{H}(\theta - h\hat{\theta^i})}{2h},
\end{equation}
where $\hat{\theta^i}$ is a unit for only the $i^\text{th}$ theta. Methods similar to this have been used in machine learning \cite{musso2020}. The random length allows for checking movement directions both within and outside the binary value for a given $\theta$.

\subsection{Warm starts}

Many commercial classical solvers improve their optimization performance by exploiting the structure of the problem. For example, warm starts are used, where the initial parameters are selected using a first guess that closely approximates a high-quality solution.
Using the non-black-box ansatz of the SQOE method, we developed strategies for warm starts. For more details on warm starting quantum optimization, see reference \cite{Egger2021warmstartingquantum}.

Each parameter - $E$, $M$, and $\vec{P}$ - poses distinct challenges. Enforcing the minimum turbine spacing $E$ in conjunction with the other constraints can be as difficult as the overall optimization problem itself. Limiting the maximum number of turbines is easy to implement as a warm start. The unfavorable locations are straightforward to implement (initializing all variables to $0$) and maintain (do not change these variables). Similar ideas to this can be found in reference \cite{QUBOpre}.

\subsection{Quantum simulation}\label{sec:QEM}
We focus on the usage of quantum-errorless, statistically noisy circuit simulations.
The Qiskit library \cite{javadi2024quantum} from IBM was used to run the simulations on the HPC system at the University of Plymouth. We did some exploratory running on IBM quantum computers, which taught us some important lessons about reducing the circuit depth used in the algorithms. We use the Gurobi classical solver \cite{gurobi} to compare with the results from the quantum solvers.

In the NISQ era of quantum computing, quantum error mitigation (QEM) techniques are a potential way to obtain reliable results from hardware with quantum errors.
Broadly, QEM techniques fall into categories such as zero-noise extrapolation (ZNE), probabilistic error cancellation (PEC) \cite{PhysRevLett.119.180509}, symmetry verification \cite{PhysRevA.98.062339}, and measurement error mitigation \cite{PhysRevA.100.052315}. Each method aims to reduce the impact of hardware-induced noise without requiring full fault tolerance, making them particularly suitable for variational or hybrid quantum algorithms. For a review on QEM techniques, see reference \cite{RevModPhys.95.045005}. We postpone the investigation of QEM for quantum solvers for the WFLO problem until later work.

\subsection{Gate counts}
The number of gates required to create an effective ansatz is a key issue in assessing a quantum algorithm’s resilience to quantum errors.
Deeper circuits are more affected by quantum noise.
The number of rotational gates and CNOT
gates for the PCE encoding are in \tref{tab:gate_counts_PCE}.
where $N$ is the number of binary variables and $n$ is the number of qubits. We find $n$ using $N=3{n\choose k}$. We used more layers for larger $k$ values to increase entanglement.
\begin{table}[H]
\centering
\begin{tabular}{lllrr}
\toprule
$N$ & $k$ & $n$ & Rotational gate count & $CNOT$ gate count \\
\midrule
16 & 1 & 6 & 24 & 10 \\
16 & 2 & 4 & 24 & 9 \\
16 & 3 & 5 & 40 & 16 \\
16 & 4 & 6 & 60 & 25 \\
49 & 1 & 17 & 68 & 32 \\
49 & 6 & 8 & 112 & 49 \\
49 & 10 & 12 & 264 & 121 \\
64 & 2 & 8 & 48 & 21 \\
64 & 3 & 7 & 56 & 24 \\
64 & 4 & 7 & 70 & 30 \\
64 & 6 & 8 & 112 & 49 \\
81 & 2 & 8 & 48 & 21 \\
81 & 3 & 7 & 56 & 24 \\
81 & 7 & 9 & 144 & 64 \\
81 & 8 & 10 & 180 & 81 \\
\bottomrule
\end{tabular}
\caption{The gate counts for the PCE-based methods across different qubit counts on which we ran experiments. $N$ is the number of binary variables. We find this as $n$ using $N=3{n\choose k}$. We used more layers for larger $k$ values to increase entanglement.}
\label{tab:gate_counts_PCE}
\end{table}

\Tref{tab:gate_counts_SQOE} reports the number of rotational gates for SQOE circuits, where the $n\leq N$ parameter can be freely selected. In later sections, we will refer to the choice of the number of qubits as the parameter $q$. The SQOE encoding uses much shallower circuits than the PCE encoding.

\begin{table}[H]
\centering
\begin{tabular}{lrr}
\toprule
$n$ & Rotational gate count & $CNOT$ gate count \\
\midrule
5 & 5 & 0 \\
6 & 6 & 0 \\
7 & 7 & 0 \\
8 & 8 & 0 \\
10 & 10 & 0 \\
15 & 15 & 0 \\
20 & 20 & 0 \\
\bottomrule
\end{tabular}
\caption{The gate counts for the SQOE-based methods across different qubit counts on which we ran experiments. $n$ is the number of qubits, which we may freely choose such that $n\leq N$. In later sections, we will refer to the choice of the number of qubits as the parameter $q$. Here, the numbers are taken as if we are using all qubits; practically, one may want to leave gaps, in which case the rotational gate count will be $\lfloor n/2\rfloor$ (to avoid crosstalk on adjacent qubits).}
\label{tab:gate_counts_SQOE}
\end{table}
\section{Results}
We examine two categories of results: the distribution of output power in \sref{sec:BPO}, and the average time to solution in \sref{sec:LTS}.

\subsection{Power output}\label{sec:BPO}
This section of results contains boxplots of the power output of 64 sampled runs. In each figure, the top plots show all solver outputs (raw results), with configurations containing an incorrect number of turbines (not equal to $M$) highlighted in red. The bottom plots display the filtered (trimmed) results after removing these invalid configurations. The Gurobi solver achieved the optimal solution for every run. The results from the Gurobi solver are shown for each system for comparison with the results from the quantum methods.

Table \ref{tab:error_results} shows how the accuracy of the results compares to Gurobi.

\begin{table}[H]
    \centering
    \begin{tabular}{lrrrrrr}
        \toprule
        Test case & $L$ & Gurobi value & Best PCE value & Best SQOE value\\
        \midrule
        Windfarm $A$ & $4$ & 4100 & 4100 & 4100 \\
        Windfarm $A$ & $7$ & 4662 & 3694 & 4662 \\
        Windfarm $A$ & $9$ & 4745 & 4126 & 4745 \\
        Windfarm $B$ & $7$ & 10250 & - & 10250 \\
        Windfarm $B$ & $9$ & 11423 & 9071 & 11423 \\
        Alltwalis windfarm & $7$ & 1000 & 998 & 1000 \\
        Alltwalis windfarm & $8$ & 987 & 980 & 987 \\
        Alltwalis windfarm & $9$ & 1006 & 1005 & 1006 \\
        \bottomrule
    \end{tabular}
    \caption{Highest power output for each method and test case that meet the number of turbines constraint. The windfarm problem details are given in full in appendix \ref{appen:TM}.}
    \label{tab:error_results}
\end{table}

\subsubsection{Windfarm $A$}
\begin{figure}[H]
  \centering
  \begin{subfigure}[t]{0.45\textwidth}
    \includegraphics[width=\linewidth]{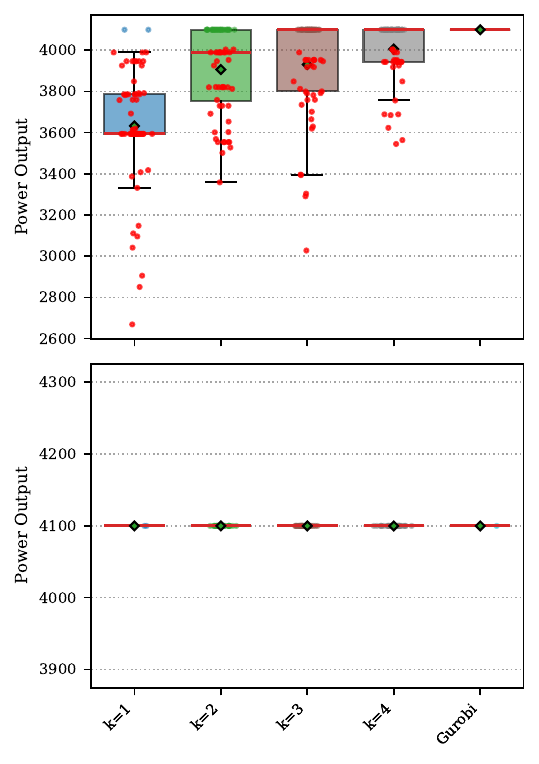}
    \caption{\footnotesize  Top: Boxplot of raw data for Windfarm $A$, $L=4$, using PCE. Bottom: Solutions with $m\neq M=16$ turbines removed.}
    \label{fig:A4P}
  \end{subfigure}
  \hfill
  \begin{subfigure}[t]{0.45\textwidth}
    \includegraphics[width=\linewidth]{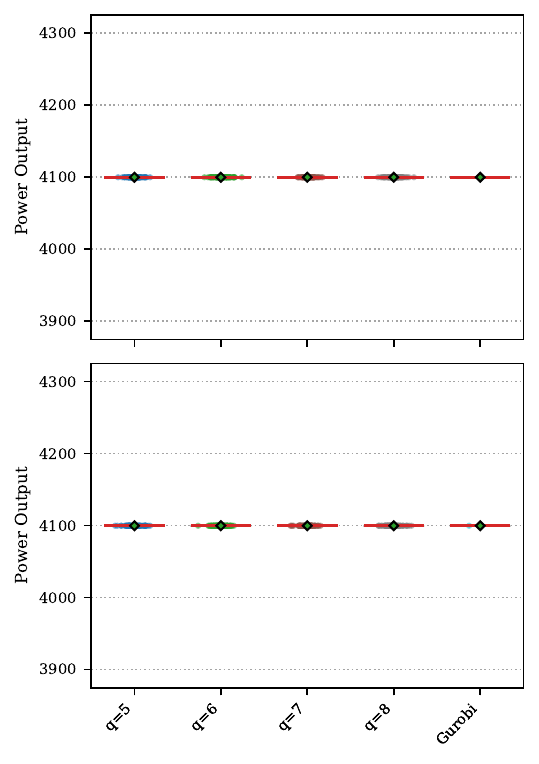}
    \caption{\footnotesize Top: Boxplot of raw data for Windfarm $A$, $L=4$, using SQOE. Bottom: Solutions with $m\neq M=16$ turbines removed.}
    \label{fig:A4S}
  \end{subfigure}
  \caption{\footnotesize Power output over sampled runs for Windfarm $A$, $L=4$ using the (a) PCE and (b) SQOE.}
\end{figure}

In \fref{fig:A4P} and \fref{fig:A4S}, we see the results for Windfarm $A$ with $L=4$. When solving the simple turbine-on-every-site solution, the SQOE showed strong performance, whereas the PCE struggled.

\begin{figure}[tbp]
  \centering
  \begin{subfigure}[t]{0.45\textwidth}
    \includegraphics[width=\linewidth]{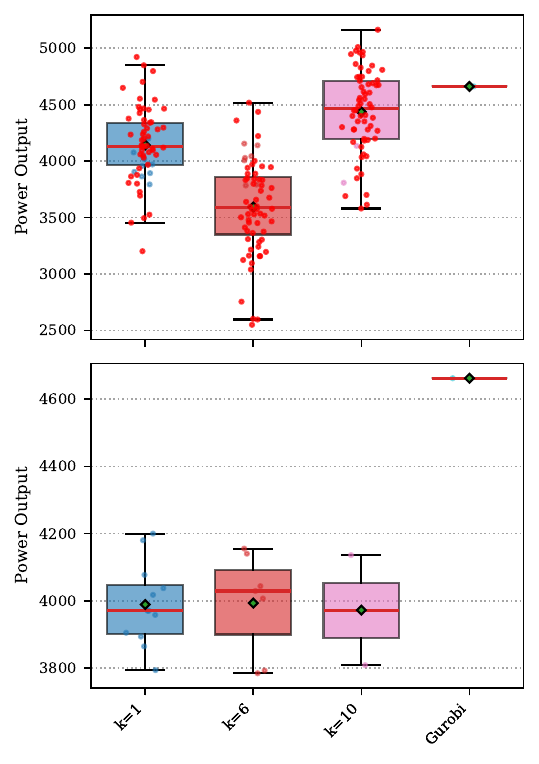}
    \caption{\footnotesize Top: Boxplot of raw data for Windfarm $A$, $L=7$, using PCE. Bottom: Solutions with $m\neq M=16$ turbines removed.}
    \label{fig:A7P}
  \end{subfigure}
  \hfill
  \begin{subfigure}[t]{0.45\textwidth}
    \includegraphics[width=\linewidth]{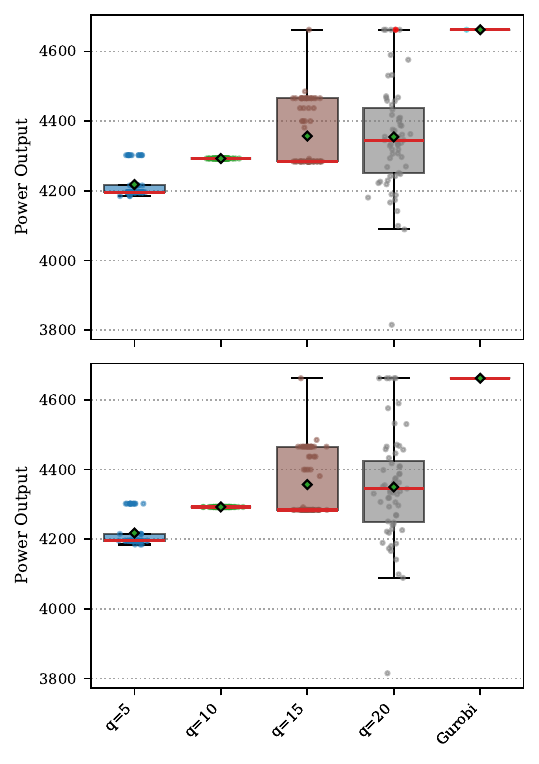}
    \caption{\footnotesize Top: Boxplot of raw data for Windfarm $A$, $L=7$, using SQOE. Bottom: Solutions with $m\neq M=16$ turbines removed.}
    \label{fig:A7S}
  \end{subfigure}
  \caption{\footnotesize Power output over sampled runs for Windfarm $A$, $L=7$ using the (a) PCE and (b) SQOE.}
\end{figure}

The second set of results, \fref{fig:A7P} and \fref{fig:A7S}, corresponds to $L=7$. The PCE did not achieve optimality, whereas the SQOE-based method successfully found the maximum power output solution. This performance improved with increasing the number of available qubits for the SQOE. The PCE frequently violated the constraints, while the SQOE was less prone to this issue. There is no clear correlation between the choice of $k$ values and the quality of the solution using the PCE.

\begin{figure}[tbp]
  \centering
  \begin{subfigure}[t]{0.45\textwidth}
    \includegraphics[width=\linewidth]{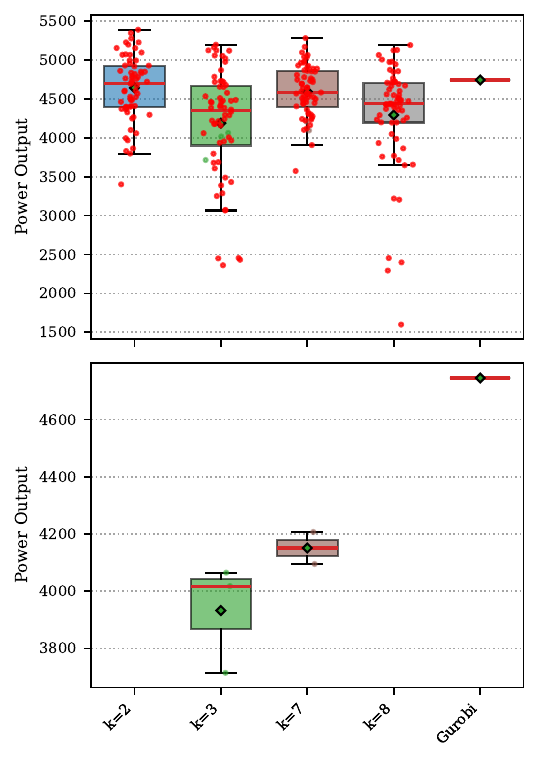}
    \caption{\footnotesize Top: Boxplot of raw data for Windfarm $A$, $L=9$, using PCE. Bottom: Solutions with $m\neq M=16$ turbines removed.}
    \label{fig:A9P}
  \end{subfigure}
  \hfill
  \begin{subfigure}[t]{0.45\textwidth}
    \includegraphics[width=\linewidth]{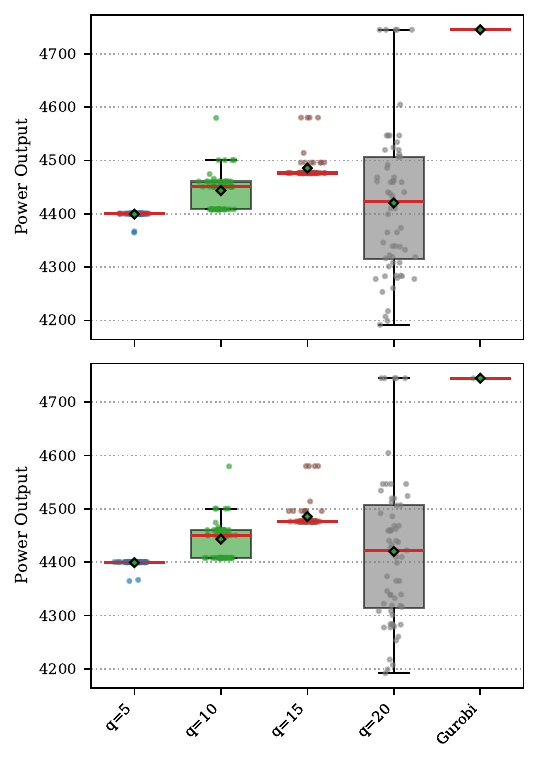}
    \caption{\footnotesize Top: Boxplot of raw data for Windfarm $A$, $L=9$, using SQOE. Bottom: Solutions with $m\neq M=16$ turbines removed.}
    \label{fig:A9S}
  \end{subfigure}
  \caption{\footnotesize Power output over sampled runs for Windfarm $A$, $L=9$ using the (a) PCE and (b) SQOE.}
\end{figure}

In \fref{fig:A9P} and \fref{fig:A9S}, we see a similar trend; the PCE struggled to stay within the constraint, while the SQOE did not violate the maximum number of turbines. Both methods produced lower average power output than Gurobi. The SQOE method was able to find optimal solutions for large $q$. 

\FloatBarrier

\subsubsection{Windfarm $B$}
\begin{figure}[H]
  \centering
  \begin{subfigure}[t]{0.45\textwidth}
    \includegraphics[width=\linewidth]{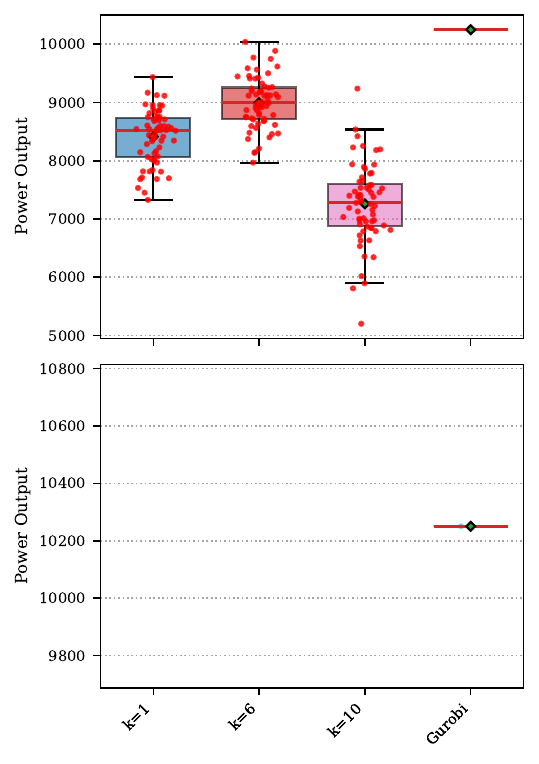}
    \caption{\footnotesize Top: Boxplot of raw data for Windfarm $B$, $L=7$, using PCE. Bottom: Solutions with $m\neq M=49$ turbines removed.}
    \label{fig:B7P}
  \end{subfigure}
  \hfill
  \begin{subfigure}[t]{0.45\textwidth}
    \includegraphics[width=\linewidth]{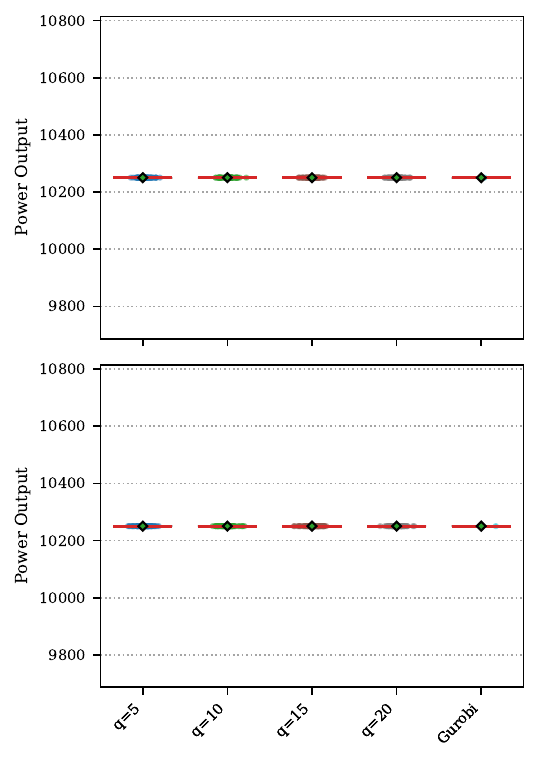}
    \caption{\footnotesize Top: Boxplot of raw data for Windfarm $B$, $L=7$, using SQOE. Bottom: Solutions with $m\neq M=49$ turbines removed.}
    \label{fig:B7S}
  \end{subfigure}
  \caption{\footnotesize Power output over sampled runs for Windfarm $B$, $L=7$ using the (a) PCE and (b) SQOE.}
\end{figure}

In \fref{fig:B7P} and \fref{fig:B7S}, we observe similar results to the previous full solution.

\begin{figure}[H]
  \centering
  \begin{subfigure}[t]{0.45\textwidth}
    \includegraphics[width=\linewidth]{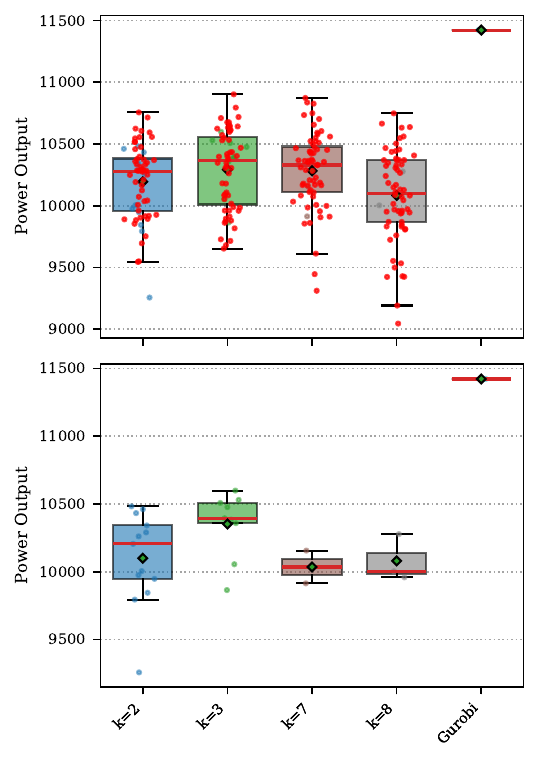}
    \caption{\footnotesize Top: Boxplot of raw data for Windfarm $B$, $L=9$, using PCE. Bottom: Solutions with $m\neq M=49$ turbines removed.}
    \label{fig:B9P}
  \end{subfigure}
  \hfill
  \begin{subfigure}[t]{0.45\textwidth}
    \includegraphics[width=\linewidth]{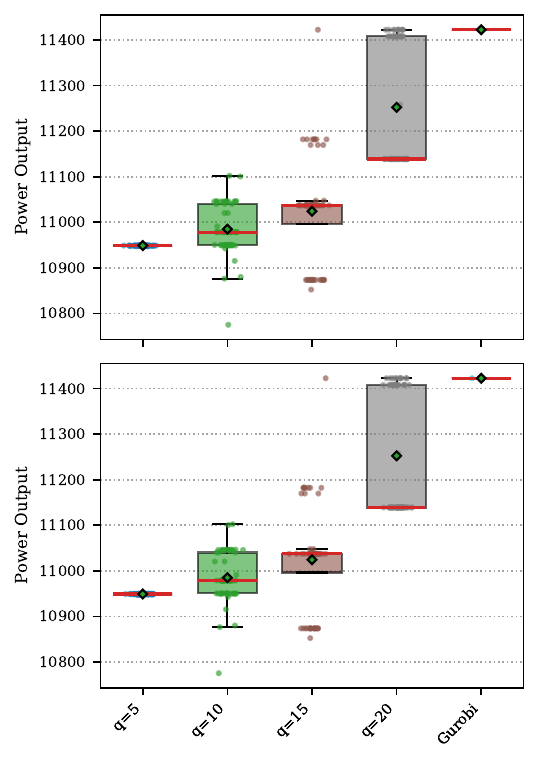}
    \caption{\footnotesize Top: Boxplot of raw data for Windfarm $B$, $L=9$, using SQOE. Bottom: Solutions with $m\neq M=49$ turbines removed.}
    \label{fig:B9S}
  \end{subfigure}
  \caption{\footnotesize Power output over sampled runs for Windfarm $B$, $L=9$ using the (a) PCE and (b) SQOE.}
\end{figure}

For the more complex case $L=9$ (shown in \fref{fig:B9P} and \fref{fig:B9S}), the PCE failed to reach the maximum power output solution. The SQOE was able to find optimality in a limited number of cases - particularly with more qubits. The results remain within a reasonable range of the best solutions found by Gurobi. As in earlier cases, no correlation was observed between parameter choice and power output for PCE-based methods. The increased scale and turbine count likely explain the convergence difficulties.

\FloatBarrier

\subsubsection{Alltwalis Windfarm}
\begin{figure}[H]
  \centering
  \begin{subfigure}[t]{0.45\textwidth}
    \includegraphics[width=\linewidth]{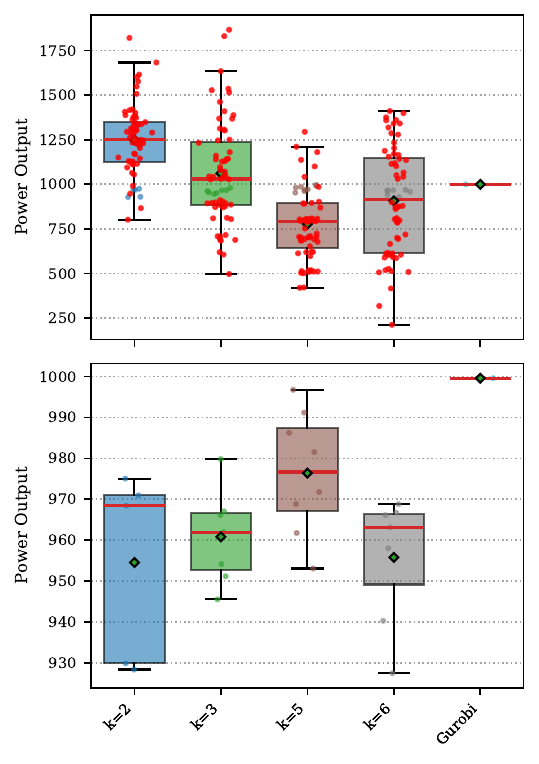}
    \caption{\footnotesize Top: Boxplot of raw data for Alltwalis windfarm, $L=7$, using PCE. Bottom: Solutions with $m\neq M=10$ turbines removed.}
    \label{fig:a7P}
  \end{subfigure}
  \hfill
  \begin{subfigure}[t]{0.45\textwidth}
    \includegraphics[width=\linewidth]{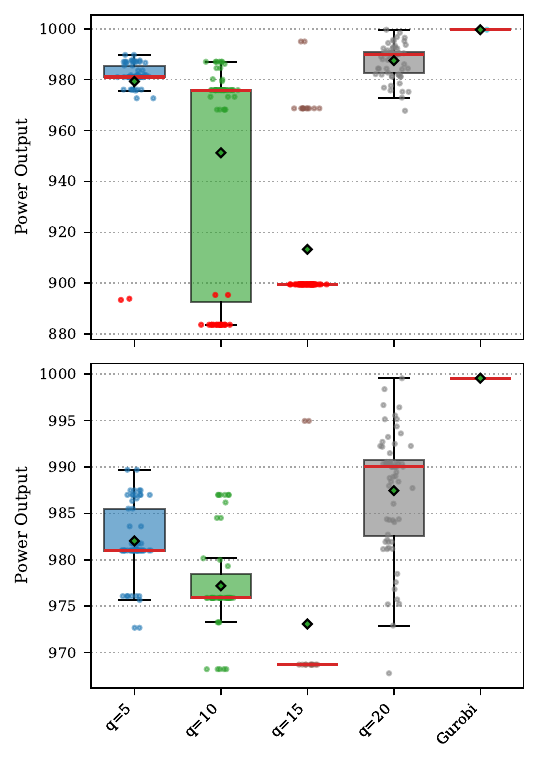}
    \caption{\footnotesize Top: Boxplot of raw data for Alltwalis windfarm, $L=7$, using SQOE. Bottom: Solutions with $m\neq M=10$ turbines removed.}
    \label{fig:a7S}
  \end{subfigure}
  \caption{\footnotesize Power output over sampled runs for Alltwalis windfarm, $L=7$ using the (a) PCE and (b) SQOE.}
\end{figure}

$L=7$ results are shown in \fref{fig:a7P} and \fref{fig:a7S}. The SQOE-based method successfully identified the optimal layout when a sufficient qubit count was employed. In contrast, the PCE method performed poorly; while some solutions were close to Gurobi in quality, many failed to satisfy the problem constraints.

\begin{figure}[tbp]
  \centering
  \begin{subfigure}[t]{0.45\textwidth}
    \includegraphics[width=\linewidth]{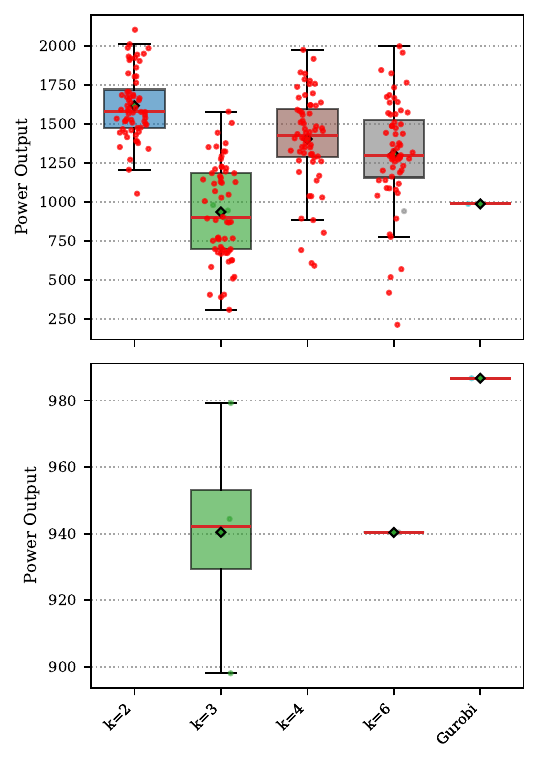}
    \caption{\footnotesize Top: Boxplot of raw data for Alltwalis windfarm, $L=8$, using PCE. Bottom: Solutions with $m\neq M=10$ turbines removed.}
    \label{fig:a8P}
  \end{subfigure}
  \hfill
  \begin{subfigure}[t]{0.45\textwidth}
    \includegraphics[width=\linewidth]{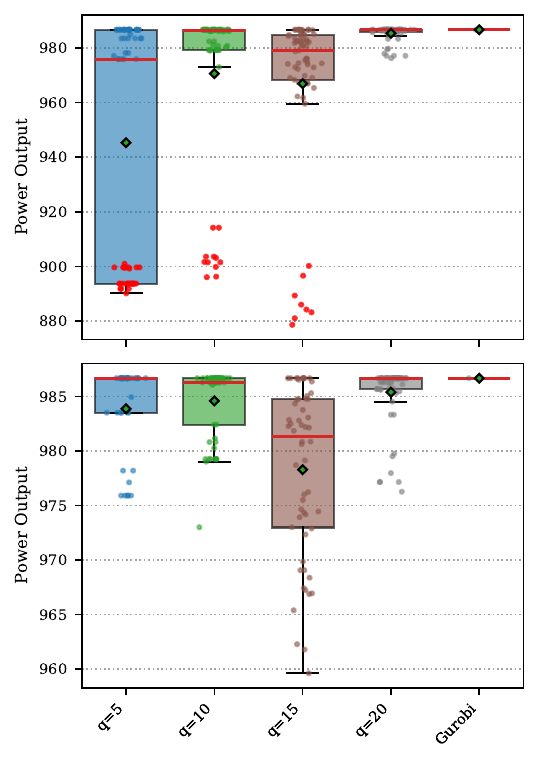}
    \caption{\footnotesize Top: Boxplot of raw data for Alltwalis windfarm, $L=8$, using SQOE. Bottom: Solutions with $m\neq M=10$ turbines removed.}
    \label{fig:a8S}
  \end{subfigure}
  \caption{\footnotesize Power output over sampled runs for Alltwalis windfarm, $L=8$ using the (a) PCE and (b) SQOE.}
\end{figure}

For the $L=8$ case, shown in \fref{fig:a8P} and \fref{fig:a8S}, the SQOE again outperformed the PCE, finding the optimal solution in all setups, and doing so with a tight spread with a large number of qubits. The PCE performed poorly, often selecting the wrong number of turbines.

\begin{figure}[H]
  \centering
  \begin{subfigure}[t]{0.45\textwidth}
    \includegraphics[width=\linewidth]{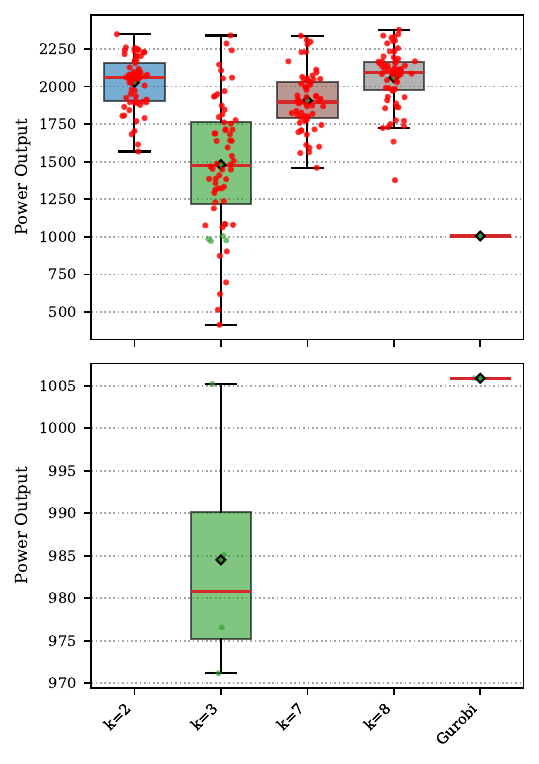}
    \caption{\footnotesize Top: Boxplot of raw data for Alltwalis windfarm, $L=9$, using PCE. Bottom: Solutions with $m\neq M=10$ turbines removed.}
    \label{fig:a9P}
  \end{subfigure}
  \hfill
  \begin{subfigure}[t]{0.45\textwidth}
    \includegraphics[width=\linewidth]{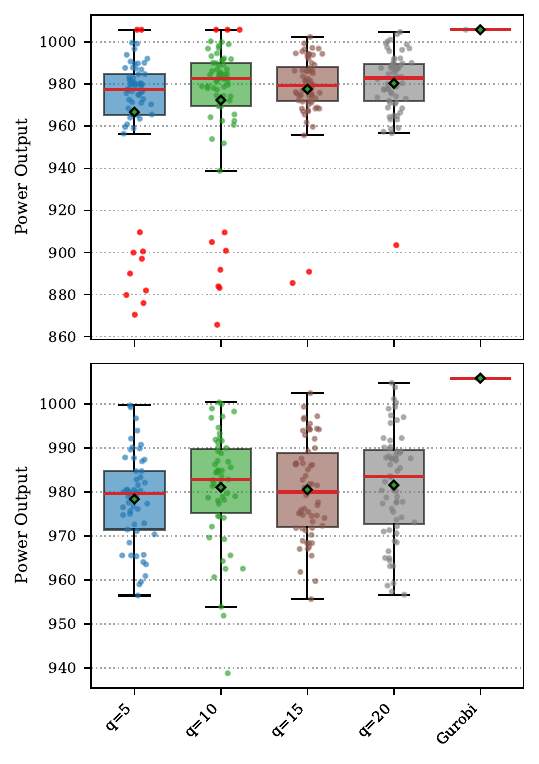}
    \caption{\footnotesize Top: Boxplot of raw data for Alltwalis windfarm, $L=9$, using SQOE. Bottom: Solutions with $m\neq M=10$ turbines removed.}
    \label{fig:a9S}
  \end{subfigure}
  \caption{\footnotesize Power output over sampled runs for Alltwalis windfarm, $L=9$ using the (a) PCE and (b) SQOE.}
\end{figure}

For $L=9$, illustrated in \fref{fig:a9P} and \fref{fig:a9S}, the trend continues: the PCE struggled to find solutions with the correct number of turbines, while the SQOE performed relatively well.

\FloatBarrier

\subsection{Time Scaling Plots}\label{sec:LTS}
This section analyzes the computational time for each method, presenting the average solve time for the samples above. Each plot includes the corresponding Gurobi solve time for reference. The labels $k_i$ or $q_i$ denote parameter groups; these are grouped as the $q_0$ being the smallest $q$ value for each scale, up to $q_3$ being the largest, and similar for $k_i$. The value of each exponent, e.g., $q_0 \sim 2.32$, indicates that the group's solve time scales as $O(N^{2.32})$,
where $N$ is the number of binary variables ($L^2$ for grid side $L$).
Linear regression fits were employed to analyze scaling trends due to limited data availability.

The full scaling results are summarized in \tref{tab:time_scaling}. The primary finding is that the quantum methods exhibited lower scaling exponents from the linear fits than Gurobi on the same problem.
\begin{table}[H]
    \centering
    \begin{tabular}{llllr}
        \toprule
        Model & Method & Variable values & Group label & Scaling exponent \\
        \midrule
        \multirow{5}{*}{Windfarm $A$} & \multirow{4}{*}{PCE} & 1, 1 & $k_0$ & 3.46 \\
         &  & 2, 6 & $k_1$ & 3.24 \\
         &  & 3, 10 & $k_2$ & 3.35 \\
        \cmidrule(lr){2-4}
        & \multirow{4}{*}{SQOE} & 5, 5, 5 & $q_0$ & 2.32 \\
         &  & 6, 10, 10 & $q_1$ & 2.3 \\
         &  & 7, 15, 15 & $q_2$ & 5.45 \\
         &  & 8, 20, 20 & $q_3$ & 5.92 \\
        \cmidrule(lr){2-4}
        & Gurobi & - & - & 6.37 \\
        \midrule
        \multirow{5}{*}{Windfarm $B$} & \multirow{4}{*}{PCE} & 1, 2 & $k_0$ & 1.86 \\
         &  & 6, 3 & $k_1$ & 0.7 \\
         &  & 10, 7 & $k_2$ & 1.96 \\
         &  & 4 & $k_3$ & 1.04 \\
        \cmidrule(lr){2-4}
        & \multirow{4}{*}{SQOE} & 5, 5 & $q_0$ & 0.29 \\
         &  & 10, 10 & $q_1$ & -0.08 \\
         &  & 15, 15 & $q_2$ & -0.04 \\
         &  & 20, 20 & $q_3$ & -0.09 \\
        \cmidrule(lr){2-4}
        & Gurobi & - & - & 6.11 \\
        \midrule
        \multirow{5}{*}{Alltwalis} & \multirow{4}{*}{PCE} & 2, 2, 2 & $k_0$ & 1.4 \\
         &  & 3, 3, 3 & $k_1$ & 2.58 \\
         &  & 5, 4, 7 & $k_2$ & 2.97 \\
        \cmidrule(lr){2-4}
        & \multirow{4}{*}{SQOE} & 5, 5, 5 & $q_0$ & 2.72 \\
         &  & 10, 10, 10 & $q_1$ & 0.86 \\
         &  & 15, 15, 15 & $q_2$ & 6.94 \\
         &  & 20, 20, 20 & $q_3$ & 3.58 \\
        \cmidrule(lr){2-4}
        & Gurobi & - & - & 13.19 \\        
        \bottomrule
    \end{tabular}
    \caption{Scaling results. The variable values are $k$ (in the choose function) for PCE and $q$ (number of qubits) for SQOE. The group labels are used in the plots below.}
    \label{tab:time_scaling}
\end{table}

\subsubsection{Windfarm $A$}
\begin{figure}[H]
  \centering
  \includegraphics[width=0.80\textwidth]{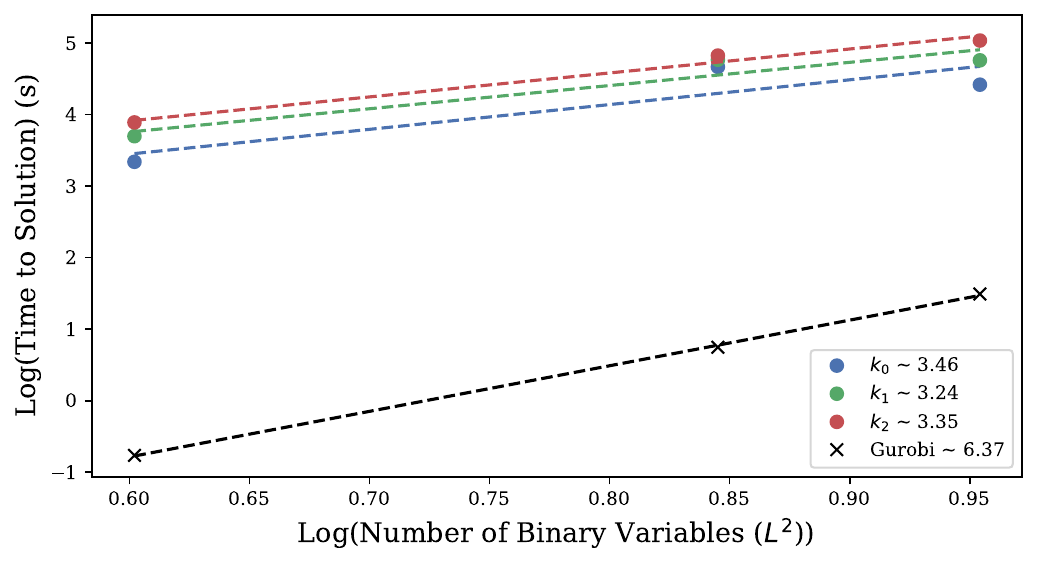}
  \caption{\footnotesize Windfarm $A$, using PCE.}
  \label{fig:timing_A_P}
\end{figure}

\begin{figure}[H]
  \centering
  \includegraphics[width=0.80\textwidth]{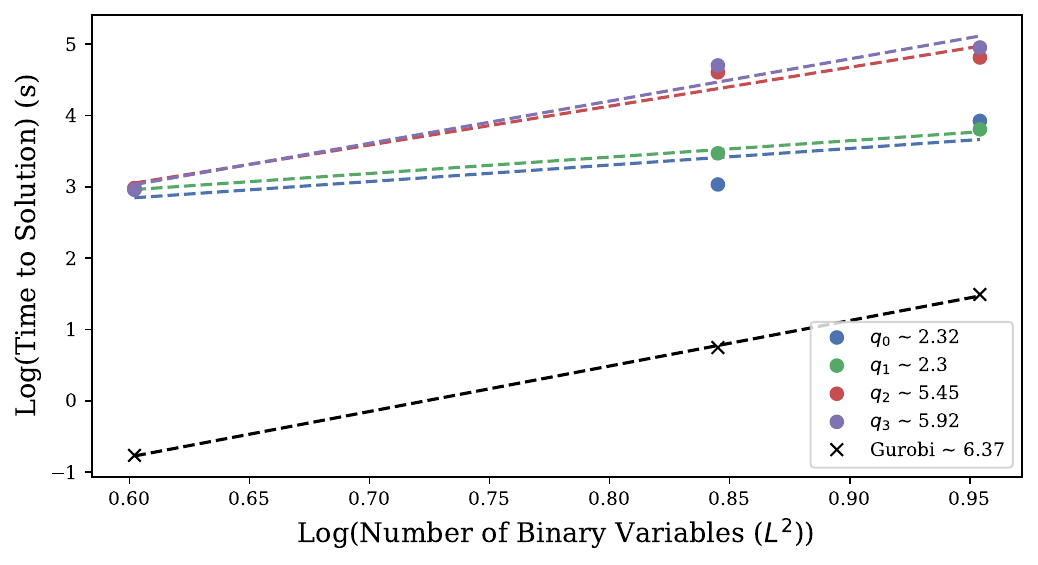}
  \caption{\footnotesize Windfarm $A$, using SQOE.}
  \label{fig:timing_A_S}
\end{figure}

Across these systems, shown in \fref{fig:timing_A_P} and \fref{fig:timing_A_S}, we see favorable scaling from the quantum methods. The SQOE's results exhibit a dependence on the chosen $q$ values, whereas the PCE-based results appear independent of $k$.

\FloatBarrier

\subsubsection{Windfarm $B$}
\begin{figure}[H]
  \centering
  \includegraphics[width=0.80\textwidth]{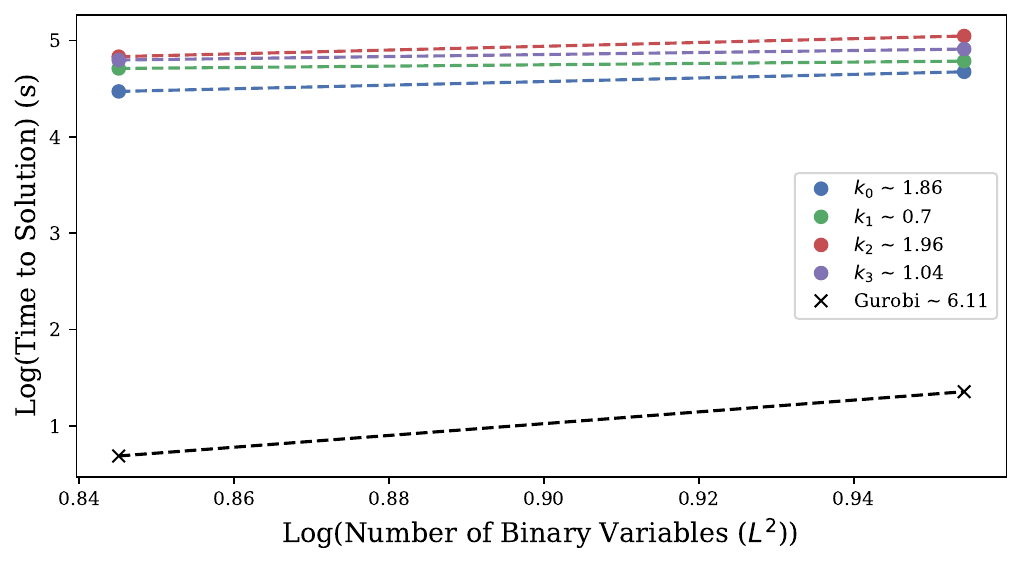}
  \caption{\footnotesize Windfarm $B$, using PCE.}
  \label{fig:timing_B_P}
\end{figure}

\begin{figure}[H]
  \centering
  \includegraphics[width=0.80\textwidth]{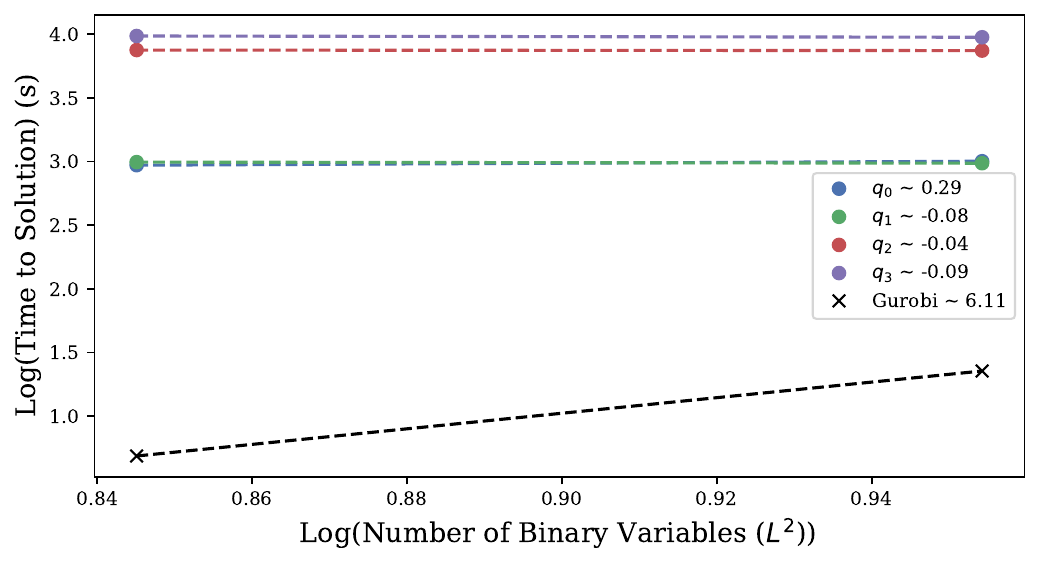}
  \caption{\footnotesize Windfarm $B$, using SQOE.}
  \label{fig:timing_B_S}
\end{figure}

\Fref{fig:timing_B_P} and \fref{fig:timing_B_S} show the scaling plots for this problem. The SQOE-based method failed to converge, which is reflected in the poor solution quality shown in the boxplots. Consequently, the computation time is dominated by the fixed overhead of using many qubits for the maximum allowed iterations. This explains the constant scaling shown by the horizontal lines. The PCE-based results also performed poorly; therefore, these results offer limited insight.

\FloatBarrier

\subsubsection{Alltwalis windfarm}
\begin{figure}[H] 
  \centering
  \includegraphics[width=0.80\textwidth]{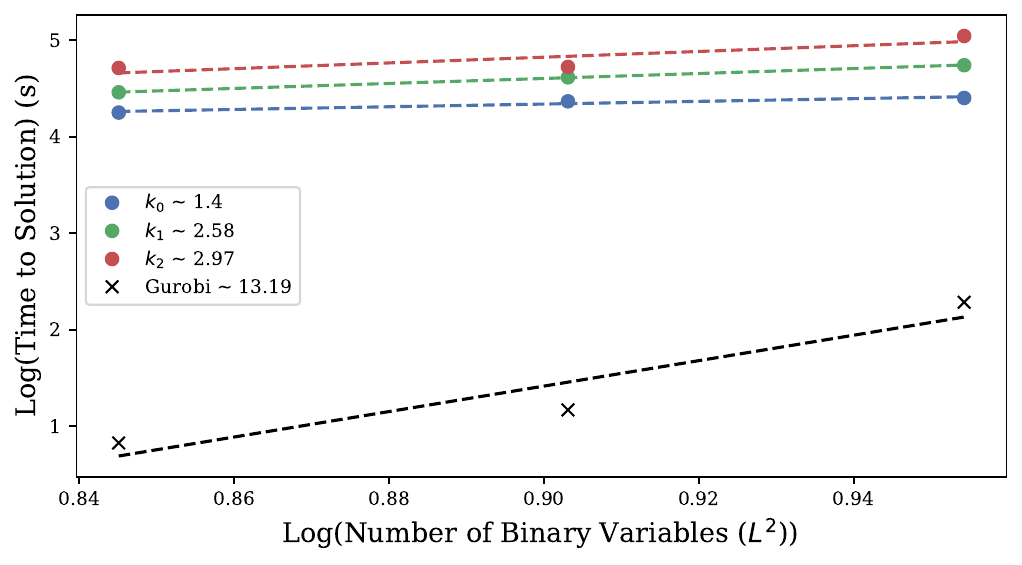}
  \caption{\footnotesize Alltwalis Windfarm, using PCE.}
  \label{fig:timing_a_P}
\end{figure}

\begin{figure}[H] 
  \centering
  \includegraphics[width=0.80\textwidth]{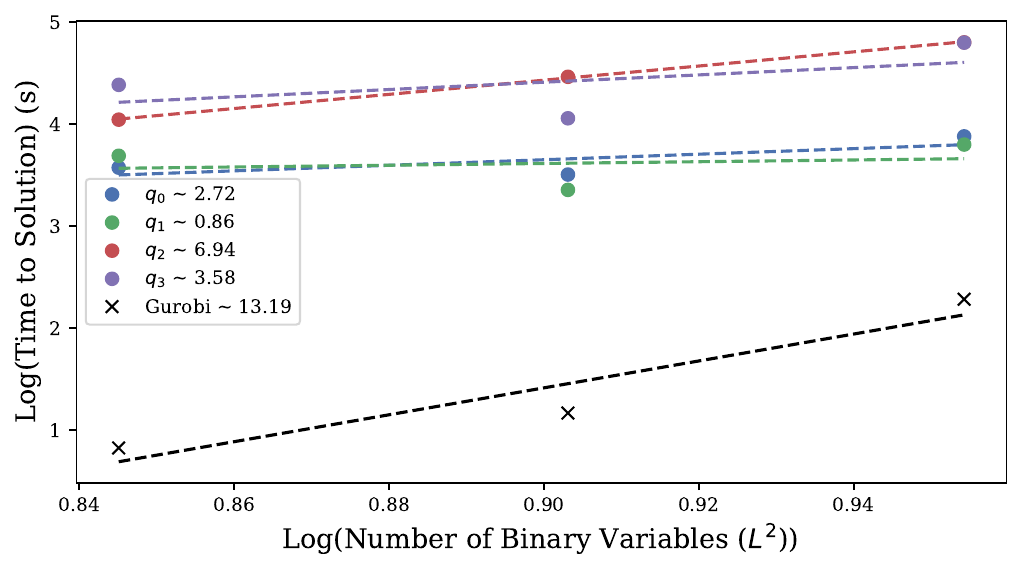}
  \caption{\footnotesize Alltwalis Windfarm, using SQOE.}
  \label{fig:timing_a_S}
\end{figure}

The results for this problem vary drastically across different choices of $k$ and $q$, as seen in \fref{fig:timing_a_P} and \ref{fig:timing_a_S}. It was on this problem that we had the best quality from the SQOE, and where it scaled the best in comparison to the Gurobi solver. This instance is the most challenging for the Gurobi solver. Consequently, the PCE method struggled to produce feasible solutions, making the scaling results difficult to interpret.

\FloatBarrier

\section{Discussion}\label{sec:CONC}

In this project, we have further developed quantum algorithms 
to solve the WFLO, building on our work in \cite{mine}. Improved wake models were used in the QUBO formalism to make the problems more physically realistic. We investigated two new encodings: PCE and SQOE, which allowed WFLOs with larger grid sizes to be solved. 


Both encodings performed fairly well with tuned hyperparameters. The SQOE method demonstrated superior performance to PCE in both power output optimization and computational time across most test cases. 
The performance of the PCE encoding was originally tested by solving the maximum cut graph problem \cite{Sciorilli2025} - the authors made use of several heuristics, and obtained good performance. It could be that additional tuning of the algorithms with the PCE encoding could improve the performance. 
The SQOE encoding achieved the best performance on the Alltwalis windfarm test case; using more qubits was advantageous, in general. The quantum-inspired approaches achieved competitive results, typically obtaining solutions within $10$-$15\%$ of the optimal values identified by Gurobi, with the SQOE-based method achieving the optimal power even in the larger systems. The quantum-based methods showed promising scaling compared to Gurobi, but overall took far longer to run.

Our study was limited to statistical noise only, with no quantum errors
included in the quantum simulators. The next step would be to include quantum errors in the simulation and test various quantum error mitigation strategies \cite{RevModPhys.95.045005}, before running on quantum computers.

It would also be useful to modify the formalism to include multiple objectives, such as minimizing the cable lengths or the number of turbines \cite{kotil2025quantum}. This could be achieved via scalarization, wherein new terms are added to the cost function, which remains a scalar value - this method comes with some limitations. The novel SQOE encoding developed in this paper may be useful to solve other optimization problems, for instance, the MaxCut problem. These results demonstrate the promising potential of quantum computing to solve optimization problems required in the area of renewable energy, such as the windfarm layout optimization problem.

\section*{Data}
Source code is publicly available (see reference \cite{Hancock2025_WindfarmLayoutOptimization}). The data that support the findings of this study are available upon reasonable request from the authors.

\section*{Acknowledgments}
We thank Davide Vadacchino for his initial contribution to this project.
Part of this project was undertaken during discussions with the Open Quantum Institute, where the WFLO problem was being developed as a use case.
This research was carried out using the computational facilities of the High Performance Computing Centre, University of Plymouth \cite{UniversityPlymouth_2025_HPC}. This project was supported by the UK National Quantum Computer Centre [NQCC200921], which is a UKRI Centre and part of the UK National Quantum Technologies Programme (NQTP). 


Grammarly \cite{Grammarly_2025} was used to help with the quality of writing throughout this article.

\bibliographystyle{elsarticle-num}
\bibliography{references_new} 

@legislation{climatechangeact2008,
	title        = {{Climate Change Act 2008}},
	year         = 2008,
	url          = {https://www.legislation.gov.uk/ukpga/2008/27/section/1},
	note         = {Accessed: 2025-11-04},
	institution  = {UK Parliament}
}

@misc{nationalgrid2023renewable,
	title        = {{How much of the {UK’s} energy is renewable?}},
	author       = {{National Grid}},
	year         = 2023,
	url          = {https://www.nationalgrid.com/stories/energy-explained/how-much-uks-energy-renewable},
	note         = {Accessed: 2025-11-04}
}

@article{wes-6-311-2021,
	title        = {{An overview of wind-energy-production prediction bias, losses, and uncertainties}},
	author       = {Lee, J. C. Y. and Fields, M. J.},
	year         = 2021,
	journal      = {Wind Energy Sci.},
	volume       = 6,
	number       = 2,
	pages        = {311--365},
	doi          = {10.5194/wes-6-311-2021}
}

@misc{bbc_wind_theft,
	title        = {{Renewable energy’s trouble with wind theft}},
	author       = {Gosden, Emily},
	year         = 2024,
	url          = {https://www.bbc.co.uk/future/article/20250506-renewable-energys-trouble-with-wind-theft},
	note         = {Accessed: 2025-11-04}
}

@article{GUIRGUIS2017279,
	title        = {{Gradient-based multidisciplinary design of wind farms with continuous-variable formulations}},
	author       = {Guirguis, David and Romero, David A. and Amon, Cristina H.},
	year         = 2017,
	journal      = {Appl. Energy},
	volume       = 197,
	pages        = {279--291},
	doi          = {10.1016/j.apenergy.2017.04.030}
}

@article{GUIRGUIS2016110,
	title        = {{Toward efficient optimization of wind farm layouts: Utilizing exact gradient information}},
	author       = {Guirguis, David and Romero, David A. and Amon, Cristina H.},
	year         = 2016,
	journal      = {Appl. Energy},
	volume       = 179,
	pages        = {110--123},
	doi          = {10.1016/j.apenergy.2016.06.101}
}

@book{jensen1983note,
	title        = {{A Note on Wind Generator Interaction}},
	author       = {Jensen, N. O.},
	year         = 1983,
	publisher    = {Ris{\o} National Laboratory},
	address      = {Roskilde, Denmark},
	series       = {Ris{\o}-M},
	number       = 2411
}

@article{jmse9121376,
	title        = {{Multi-Objective Optimisation of the Benchmark Wind Farm Layout Problem}},
	author       = {Manikowski, Pawel L. and Walker, David J. and Craven, Matthew J.},
	year         = 2021,
	journal      = {J. Mar. Sci. Eng.},
	volume       = 9,
	number       = 12,
	doi          = {10.3390/jmse9121376},
	article      = 1376
}

@article{LI2022100307,
	title        = {{Discrete complex-valued code pathfinder algorithm for wind farm layout optimization problem}},
	author       = {Li, Ning and Zhou, Yongquan and Luo, Qifang and Huang, Huajuan},
	year         = 2022,
	journal      = {Energy Convers. Manage.: X},
	volume       = 16,
	pages        = 100307,
	doi          = {10.1016/j.ecmx.2022.100307}
}

@misc{dong2024reinforcementlearningenhancedgeneticalgorithm,
	title        = {{Reinforcement learning-enhanced genetic algorithm for wind farm layout optimization}},
	author       = {Dong, Guodan and Qin, Jianhua and Wu, Chutian and Xu, Chang and Yang, Xiaolei},
	year         = 2024,
	doi          = {10.2139/ssrn.5038903},
	eprint       = {2412.06803},
	archiveprefix = {arXiv},
	primaryclass = {cs.NE}
}

@article{RAMLI2023101016,
	title        = {{Wind farm layout optimization using a multi-objective electric charged particles optimization and a variable reduction approach}},
	author       = {Ramli, Makbul A.M. and Bouchekara, Houssem R.E.H. and Milyani, Ahmad H.},
	year         = 2023,
	journal      = {Energy Strategy Rev.},
	volume       = 45,
	pages        = 101016,
	doi          = {10.1016/j.esr.2022.101016}
}

@misc{pedersen2025integrated,
	title        = {{Integrated Wind Farm Design: Optimizing Turbine Placement and Cable Routing with Wake Effects}},
	author       = {Pedersen, Jaap and Lindner, Niels and Rehfeldt, Daniel and Koch, Thorsten},
	year         = 2025,
	doi          = {10.48550/arXiv.2501.07203},
	eprint       = {2501.07203},
	archiveprefix = {arXiv},
	primaryclass = {math.OC}
}

@article{QUBOsurvey,
	title        = {{The unconstrained binary quadratic programming problem: A survey}},
	author       = {Kochenberger, G. and Hao, J.-K. and Glover, F. and Lewis, M. and Lü, Z. and Wang, H. and Wang, Y.},
	year         = 2014,
	journal      = {J. Comb. Optim.},
	volume       = 28,
	number       = 1,
	pages        = {58--81},
	doi          = {10.1007/s10878-014-9734-0}
}

@article{MOSETTI1994105,
	title        = {{Optimization of wind turbine positioning in large windfarms by means of a genetic algorithm}},
	author       = {Mosetti, G. and Poloni, C. and Diviacco, B.},
	year         = 1994,
	journal      = {J. Wind Eng. Ind. Aerodyn.},
	volume       = 51,
	number       = 1,
	pages        = {105--116},
	doi          = {10.1016/0167-6105(94)90080-9}
}

@article{senderovich2022exploiting,
	title        = {{Exploiting Hardware and Software Advances for Quadratic Models of Wind Farm Layout Optimization}},
	author       = {Senderovich, Arik and Zhang, Jiachen and Cohen, Eldan and Beck, J Christopher},
	year         = 2022,
	journal      = {IEEE Access},
	volume       = 10,
	pages        = {78044--78055},
	doi          = {10.1109/ACCESS.2022.3193143}
}

@inproceedings{SSproceedings,
	title        = {{Solving Wind Farm Layout Optimization with Mixed Integer Programming and Constraint Programming}},
	author       = {Zhang, Peter and Romero, David and Beck, J. Christopher and Amon, Cristina},
	year         = 2013,
	booktitle    = {Integration of AI and OR Techniques in Constraint Programming for Combinatorial Optimization Problems},
	pages        = {284--299},
	doi          = {10.1007/978-3-642-38171-3_19},
	isbn         = {978-3-642-38170-6}
}

@inproceedings{Donovan2005WindFO,
	title        = {{An Improved Mixed Integer Programming Model for Wind Farm Layout Optimisation}},
	author       = {Donovan, Stuart},
	year         = 2006,
	booktitle    = {Proc. 41st Annu. Conf. Oper. Res. Soc. New Zealand},
	pages        = {143--152},
	url          = {https://www.researchgate.net/publication/228814242}
}

@inproceedings{3a81166868144671af170595fd17b8f6,
	title        = {{A Simple Model for Cluster Efficiency}},
	author       = {Katic, I. and H{\o}jstrup, J. and Jensen, N.O.},
	year         = 1987,
	booktitle    = {EWEC'86. Proceedings. Vol. 1},
	publisher    = {A. Raguzzi},
	pages        = {407--410},
	editor       = {Palz, W. and Sesto, E.}
}

@article{Sciorilli2025,
	title        = {{Towards large-scale quantum optimization solvers with few qubits}},
	author       = {Sciorilli, Marco and Borges, Lucas and Patti, Taylor L. and Garc{\'i}a-Mart{\'i}n, Diego and Camilo, Giancarlo and Anandkumar, Anima and Aolita, Leandro},
	year         = 2025,
	journal      = {Nat. Commun.},
	volume       = 16,
	number       = 1,
	pages        = 476,
	doi          = {10.1038/s41467-024-55346-z}
}

@article{PRXQuantum.3.020301,
	title        = {{Quantum Crosstalk Analysis for Simultaneous Gate Operations on Superconducting Qubits}},
	author       = {Zhao, Peng and Linghu, Kehuan and Li, Zhiyuan and Xu, Peng and Wang, Ruixia and Xue, Guangming and Jin, Yirong and Yu, Haifeng},
	year         = 2022,
	journal      = {PRX Quantum},
	volume       = 3,
	pages        = {020301},
	doi          = {10.1103/PRXQuantum.3.020301}
}

@article{COBYLA,
	title        = {{Efficient strategies for constrained black-box optimization by intrinsically linear approximation (CBOILA)}},
	author       = {Liu, Chengyang and Wan, Zhiqiang and Li, Xuewu and Liu, Dianzi},
	year         = 2022,
	journal      = {Eng. Comput.},
	volume       = 38,
	number       = 1,
	pages        = {401--414},
	doi          = {10.1007/s00366-020-01160-2}
}

@article{10.1214/aoms/1177729586,
	title        = {{A Stochastic Approximation Method}},
	author       = {Robbins, Herbert and Monro, Sutton},
	year         = 1951,
	journal      = {Ann. Math. Stat.},
	volume       = 22,
	number       = 3,
	pages        = {400--407},
	doi          = {10.1214/aoms/1177729586}
}

@incollection{LeCun1998Efficient,
	title        = {{Efficient BackProp}},
	author       = {LeCun, Yann and Bottou, L{\'e}on and Orr, Genevieve B. and M{\"u}ller, Klaus-Robert},
	year         = 1998,
	booktitle    = {Neural Networks: Tricks of the Trade},
	publisher    = {Springer},
	pages        = {9--50},
	doi          = {10.1007/3-540-49430-8_2},
	editor       = {Orr, Genevieve B. and M{\"u}ller, Klaus-Robert}
}

@article{Egger2021warmstartingquantum,
	title        = {{Warm-starting quantum optimization}},
	author       = {Egger, Daniel J. and Mare{\v{c}}ek, Jakub and Woerner, Stefan},
	year         = 2021,
	journal      = {Quantum},
	volume       = 5,
	pages        = 479,
	doi          = {10.22331/q-2021-06-17-479}
}

@article{QUBOpre,
	title        = {{Quadratic unconstrained binary optimization problem preprocessing: Theory and empirical analysis}},
	author       = {Lewis, Mark and Glover, Fred},
	year         = 2017,
	journal      = {Networks},
	volume       = 70,
	number       = 2,
	pages        = {79--97},
	doi          = {10.1002/net.21751}
}

@article{RevModPhys.95.045005,
	title        = {{Quantum error mitigation}},
	author       = {Cai, Zhenyu and Babbush, Ryan and Benjamin, Simon C. and Endo, Suguru and Huggins, William J. and Li, Ying and McClean, Jarrod R. and O'Brien, Thomas E.},
	year         = 2023,
	journal      = {Rev. Mod. Phys.},
	volume       = 95,
	pages        = {045005},
	doi          = {10.1103/RevModPhys.95.045005}
}

@article{PhysRevLett.119.180509,
	title        = {{Error Mitigation for Short-Depth Quantum Circuits}},
	author       = {Temme, Kristan and Bravyi, Sergey and Gambetta, Jay M.},
	year         = 2017,
	journal      = {Phys. Rev. Lett.},
	volume       = 119,
	pages        = 180509,
	doi          = {10.1103/PhysRevLett.119.180509}
}

@article{PhysRevA.98.062339,
	title        = {{Low-cost error mitigation by symmetry verification}},
	author       = {Bonet-Monroig, X. and Sagastizabal, R. and Singh, M. and O'Brien, T. E.},
	year         = 2018,
	journal      = {Phys. Rev. A},
	volume       = 98,
	pages        = {062339},
	doi          = {10.1103/PhysRevA.98.062339}
}

@article{PhysRevA.100.052315,
	title        = {{Detector tomography on IBM quantum computers and mitigation of an imperfect measurement}},
	author       = {Chen, Yanzhu and Farahzad, Maziar and Yoo, Shinjae and Wei, Tzu-Chieh},
	year         = 2019,
	journal      = {Phys. Rev. A},
	volume       = 100,
	pages        = {052315},
	doi          = {10.1103/PhysRevA.100.052315}
}

@article{rod2016,
	title        = {{Multi-objective optimization of wind farm layouts – Complexity, constraint handling and scalability}},
	author       = {Rodrigues, Silvio Cesar and Bauer, P. and Bosman, Peter},
	year         = 2016,
	journal      = {Renew. Sustain. Energy Rev.},
	volume       = 65,
	pages        = {587--609},
	doi          = {10.1016/j.rser.2016.07.021}
}

@misc{thewindpower_alltwalis_2025,
	title        = {{Alltwalis (United-Kingdom)}},
	author       = {{The Wind Power}},
	year         = 2025,
	url          = {https://www.thewindpower.net/windfarm_en_1613_alltwalis.php},
	note         = {Accessed: 2025-11-04}
}

@inproceedings{NSWD,
	title        = {{Modelling of offshore wind resources. Comparison of a meso-scale model and measurements from {FINO} 1 and {N}orth {S}ea oil rigs}},
	author       = {Berge, Erik and Byrkjedal, {{\O{}}}yvind and Ydersbond, Yngve and Kindler, Detlef},
	year         = 2009,
	booktitle    = {EWEC 2009},
	pages        = {1--8}
}

@misc{statkraft_alltwalis,
	title        = {{Alltwalis Hydropower Scheme}},
	author       = {{Statkraft UK}},
	year         = 2025,
	url          = {https://www.statkraft.co.uk/about-statkraft-uk/where-we-operate/Locations/alltwalis/},
	note         = {Accessed: 2025-11-04}
}

@misc{GlobalWindAtlas2025,
	title        = {{Global Wind Atlas}},
	author       = {{Technical University of Denmark (DTU) and World Bank Group}},
	year         = 2025,
	url          = {https://globalwindatlas.info/en},
	note         = {Accessed: 2025-11-04}
}

@misc{siemans,
	title        = {{Siemens SWT-2.3-93}},
	author       = {{The Wind Power}},
	year         = 2025,
	url          = {https://www.thewindpower.net/turbine_en_22_siemens_swt-2.3-93.php},
	note         = {Accessed: 2025-11-04}
}

@article{wes-5-1551-2020,
	title        = {{Integrated wind farm layout and control optimization}},
	author       = {Pedersen, M. M. and Larsen, G. C.},
	year         = 2020,
	journal      = {Wind Energy Sci.},
	volume       = 5,
	number       = 4,
	pages        = {1551--1566},
	doi          = {10.5194/wes-5-1551-2020}
}

@book{burton2011wind,
	title        = {{Wind Energy Handbook}},
	author       = {Burton, Tony and Sharpe, David and Jenkins, Nick and Bossanyi, Ervin},
	year         = 2011,
	publisher    = {John Wiley \& Sons},
	doi          = {10.1002/9781119992714},
	isbn         = 9780470699751,
	edition      = {2nd}
}

@misc{musso2020,
	title        = {{Stochastic gradient descent with random learning rate}},
	author       = {Musso, Daniele},
	year         = 2020,
	eprint       = {2003.06926},
	archiveprefix = {arXiv},
	primaryclass = {cs.LG}
}

@article{mine,
	title        = {{Investigating Techniques to Optimise the Layout of Turbines in a Windfarm Using a Quantum Computer}},
	author       = {Hancock, James and Craven, Matthew and McNeile, Craig and Vadacchino, Davide},
	year         = 2025,
	journal      = {J. Quantum Comput.},
	volume       = 7,
	number       = 1,
	pages        = {55--79},
	doi          = {10.32604/jqc.2025.068127}
}

@article{MANIKOWSKI2025112879,
	title        = {{Many-objective optimisation of offshore wind farms}},
	author       = {Manikowski, Pawel L. and Craven, Matthew J. and Walker, David J.},
	year         = 2025,
	journal      = {Appl. Soft Comput.},
	volume       = 173,
	pages        = 112879,
	doi          = {10.1016/j.asoc.2025.112879}
}

@misc{Hancock2025_WindfarmLayoutOptimization,
	title        = {{WindfarmLayoutOptimization$\_$2 - {GitHub} repository}},
	author       = {Hancock, James},
	year         = 2025,
	url          = {{https://github.com/jamesshancock/WindfarmLayoutOptimization_2/tree/main}},
	note         = {Accessed: 2025-11-04}
}

@misc{UniversityPlymouth_2025_HPC,
	title        = {{Lovelace System – High Performance Computing (HPC) facility}},
	author       = {{University of Plymouth}},
	year         = 2025,
	url          = {https://www.plymouth.ac.uk/facilities/high-performance-computing},
	note         = {Accessed: 2025-11-04}
}

@software{Grammarly_2025,
	title        = {{Grammarly}},
	author       = {{Grammarly Inc.}},
	year         = 2025,
	url          = {https://www.grammarly.com/},
	note         = {Accessed: 2025-11-04}
}

@misc{nqcc2024,
    author       = {{National Quantum Computing Centre}},
    title        = {{National Quantum Computing Centre (NQCC)}},
    year         = {2024},
    url          = {https://www.nqcc.ac.uk/},
    note         = {Accessed: 2025-11-04}
}

@article{kagemoto2024possible,
  title={Possible application of quantum computing in the field of ocean engineering: optimization of an offshore wind farm layout with the Ising model},
  author={Kagemoto, Hiroshi},
  journal={Journal of Ocean Engineering and Marine Energy},
  pages={1--10},
  year={2024},
  publisher={Springer}
}

@article{john2008implementation,
  title={Implementation of warm-start strategies in interior-point methods for linear programming in fixed dimension},
  author={John, Elizabeth and Y{\i}ld{\i}r{\i}m, E Alper},
  journal={Computational Optimization and Applications},
  volume={41},
  number={2},
  pages={151--183},
  year={2008},
  publisher={Springer}
}

@article{truger2024warm,
  title={Warm-starting and quantum computing: A systematic mapping study},
  author={Truger, Felix and Barzen, Johanna and Bechtold, Marvin and Beisel, Martin and Leymann, Frank and Mandl, Alexander and Yussupov, Vladimir},
  journal={ACM Computing Surveys},
  volume={56},
  number={9},
  pages={1--31},
  year={2024},
  publisher={ACM New York, NY}
}

@article{kotil2025quantum,
  title={Quantum approximate multi-objective optimization},
  author={Kotil, Ayse and Pelofske, Elijah and Riedm{\"u}ller, Stephanie and Egger, Daniel J and Eidenbenz, Stephan and Koch, Thorsten and Woerner, Stefan},
  journal={Nature Computational Science},
  pages={1--10},
  year={2025},
  publisher={Nature Publishing Group US New York}
}

@Inbook{Cela2022,
author="{\c{C}}ela, Eranda
and Punnen, Abraham P.",
editor="Punnen, Abraham P.",
title="Complexity and Polynomially Solvable Special Cases of QUBO",
bookTitle="The Quadratic Unconstrained Binary Optimization Problem: Theory, Algorithms, and Applications",
year="2022",
publisher="Springer International Publishing",
address="Cham",
pages="57--95",
isbn="978-3-031-04520-2",
doi="10.1007/978-3-031-04520-2_3",
url="https://doi.org/10.1007/978-3-031-04520-2_3"
}

@inproceedings{liu2025framework,
  title={A Framework for Automatically Setting Multiple Penalty Weights in Quadratic Unconstrained Binary Optimization},
  author={Liu, Jiajie and Moraglio, Alberto},
  booktitle={Proceedings of the Genetic and Evolutionary Computation Conference Companion},
  pages={575--578},
  year={2025}
}

@article{javadi2024quantum,
  title={Quantum computing with Qiskit},
  author={Javadi-Abhari, Ali and Treinish, Matthew and Krsulich, Kevin and Wood, Christopher J and Lishman, Jake and Gacon, Julien and Martiel, Simon and Nation, Paul D and Bishop, Lev S and Cross, Andrew W and others},
  journal={arXiv preprint arXiv:2405.08810},
  year={2024}
}

@article{nigro2025leveraging,
  title={Leveraging Quantum Annealing for Layout Optimization},
  author={Nigro, Luca and Sala, Simone and Amendola, Alfonso and Prati, Enrico},
  journal={Advanced Quantum Technologies},
  volume={8},
  number={11},
  pages={e00358},
  year={2025},
  publisher={Wiley Online Library}
}

@misc{gurobi,
  author = {{Gurobi Optimization, LLC}},
  title = {{Gurobi Optimizer Reference Manual}},
  year = 2024,
  url = "https://www.gurobi.com"
}

@article{Abbas:2023agz,
    author = "Abbas, Amira and others",
    title = "{Challenges and opportunities in quantum optimization}",
    eprint = "2312.02279",
    archivePrefix = "arXiv",
    primaryClass = "quant-ph",
    doi = "10.1038/s42254-024-00770-9",
    journal = "Nature Rev. Phys.",
    volume = "6",
    number = "12",
    pages = "718--735",
    year = "2024"
}

@article{Carmo:2025tqt,
    author = "Carmo, Rafael S. do and Reis, Renato Gomes dos and Silva, Samuel Fernando F. and Arruda, Luiz Gustavo E. and Fanchini, Felipe F.",
    title = "{Warm-Starting PCE for Traveling Salesman Problem}",
    eprint = "2509.14414",
    archivePrefix = "arXiv",
    primaryClass = "quant-ph",
    month = "9",
    year = "2025"
}
\appendix
\section*{Appendices}
\section{Wake Patterns}\label{appen:Wakes}
We visualize wake patterns by calculating reduced wind speeds at all sites in the presence of turbines.  The results are shown as high-resolution heatmaps. The physical features are those from Windfarm A in \tref{tab:summary} in appendix \ref{appen:TM}.

\begin{figure}[H]
  \centering
  \begin{subfigure}[t]{0.48\textwidth}
    \centering
    \includegraphics[width=\linewidth]{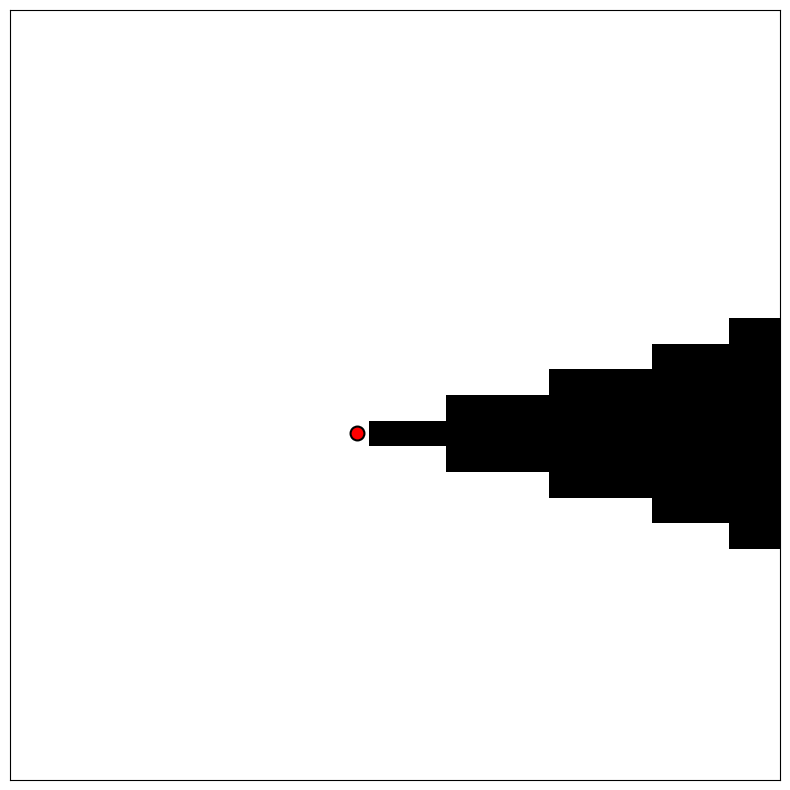}
    \caption{Base wake of effect, with a simple wind regime $D = \{\{0,v,1\}\}$, to visualize wake patterns. One turbine has been placed.}
    \label{fig:blackwakes1}
  \end{subfigure}%
  \hfill
  \begin{subfigure}[t]{0.48\textwidth}
    \centering
    \includegraphics[width=\linewidth]{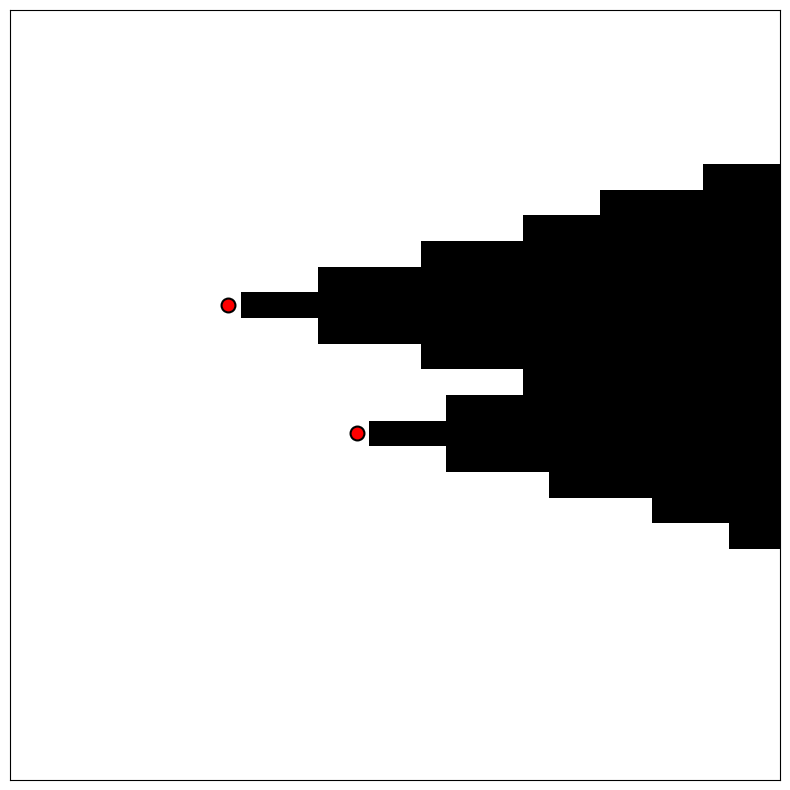}
    \caption{Base wake of effect, with a simple wind regime $D = \{\{0,v,1\}\}$, to visualize wake patterns. Two turbines have been placed.}
    \label{fig:blackwakes2}
  \end{subfigure}
  \caption{Visualization of base wake effects on the grid with one and two turbines.}
\end{figure}

\begin{figure}[H]
  \centering
  \begin{subfigure}[t]{0.48\textwidth}
    \centering
    \includegraphics[width=\linewidth]{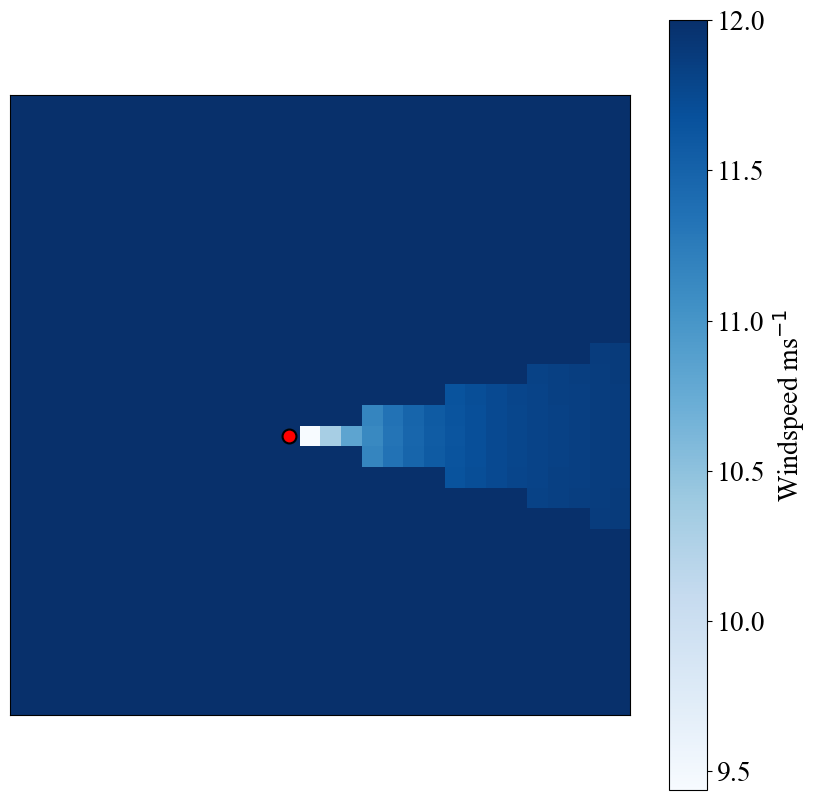}
    \caption{Heatmap of windspeeds on a $30\times30$ grid with wind regime $D = \{\{0,12\text{ms}^{-1},1\}\}$. One turbine has been placed.}
    \label{fig:calcwakes1}
  \end{subfigure}%
  \hfill
  \begin{subfigure}[t]{0.48\textwidth}
    \centering
    \includegraphics[width=\linewidth]{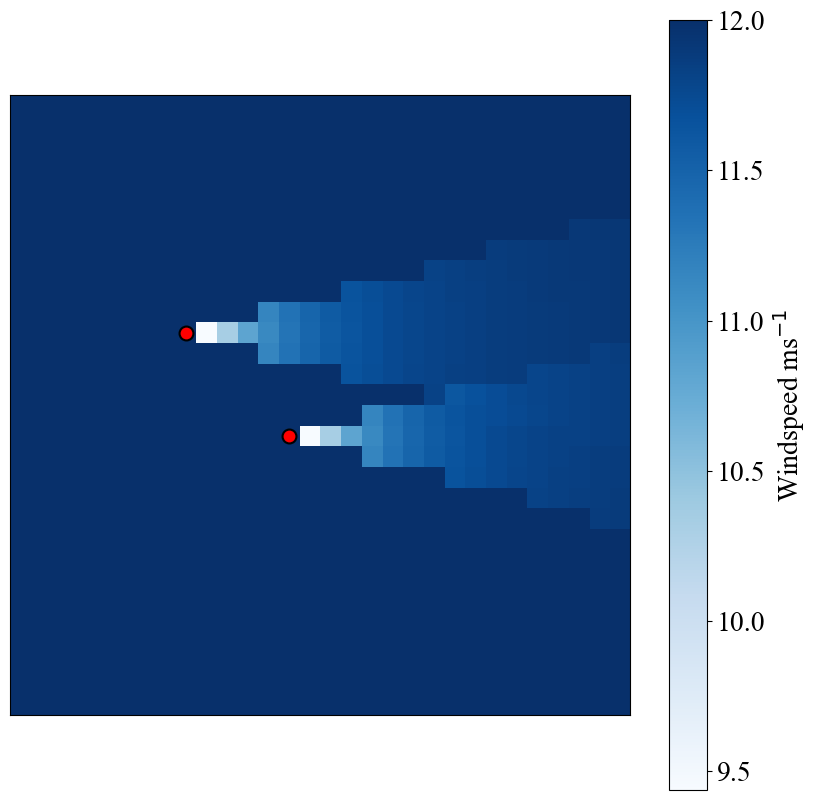}
    \caption{Heatmap of windspeeds on a $30\times30$ grid with wind regime $D = \{\{0,12\text{ms}^{-1},1\}\}$. Two turbines have been placed.}
    \label{fig:calcwakes2}
  \end{subfigure}
  \caption{Heatmaps showing wake dynamics on the grid for one and two turbines.}
\end{figure}

The base wake pattern is illustrated using a simple wind regime defined as $D = \{\{0,v,1\}\}$. Figures \ref{fig:blackwakes1} and \ref{fig:blackwakes2} display the resulting wakes for configurations of one and two turbines on a grid with $L=30$. We can introduce a windspeed, $v= 12\text{ms}^{-1}$, and calculate the windspeeds at all sites - this is shown in \fref{fig:calcwakes1} and \fref{fig:calcwakes2}. 

\begin{figure}[h!]
  \centering
  \begin{subfigure}[t]{0.48\textwidth}
    \centering
    \includegraphics[width=\linewidth]{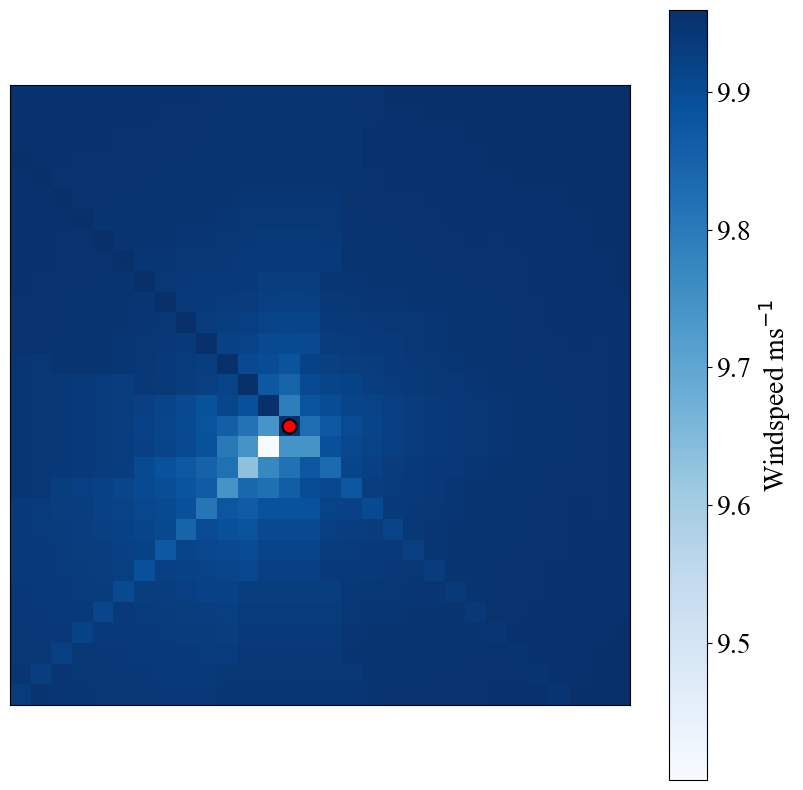}
    \caption{Heatmap of windspeeds on a $30\times30$ grid with the wind regime from \tref{tab:literature_windregime}, to visualize wake dynamics. One turbine has been placed.}
    \label{fig:compwakes1}
  \end{subfigure}%
  \hfill
  \begin{subfigure}[t]{0.48\textwidth}
    \centering
    \includegraphics[width=\linewidth]{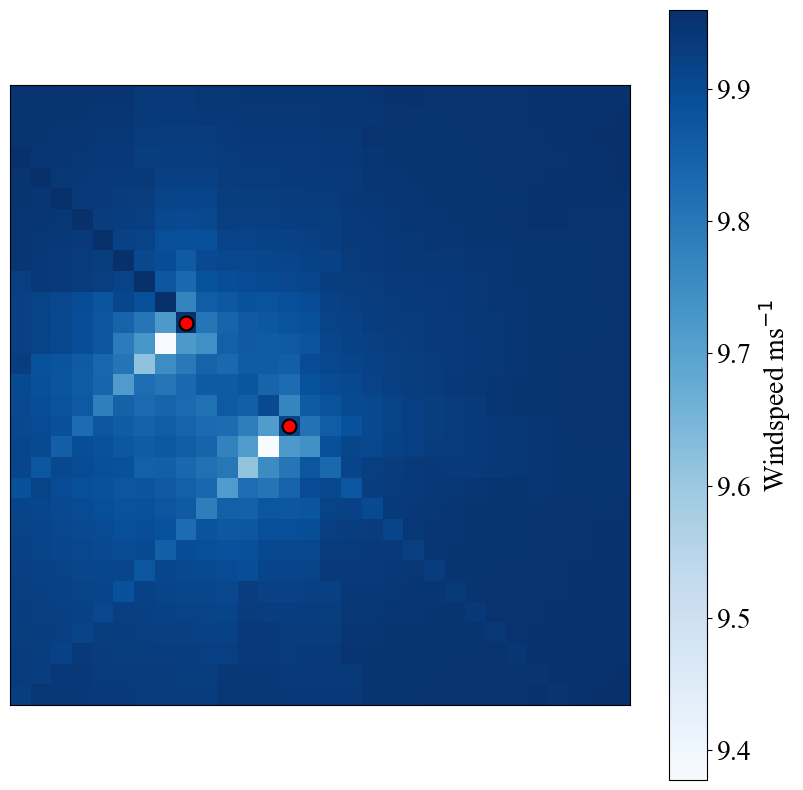}
    \caption{Heatmap of windspeeds on a $30\times30$ grid with the wind regime from \tref{tab:literature_windregime}, to visualize wake dynamics. Two turbines have been placed.}
    \label{fig:compwakes2}
  \end{subfigure}
  \caption{Heatmaps showing wake dynamics using the more complex North Sea wind regime for one and two turbines.}
\end{figure}

More complex wind regimes, including the North Sea regime (see \tref{tab:literature_windregime} in appendix \ref{appen:TM}), are calculated using a similar methodology. Examples with this wind regime, where one and two turbines are placed, are shown in \fref{fig:compwakes1} and \fref{fig:compwakes2}, respectively. 

\section{Test Models}\label{appen:TM}
While in \sref{sec:WFLO} we constructed our model in terms of grid sizes, when building more realistic models, it is advantageous to reverse our view to resolutions. The resolution of our windfarms is often given in terms of the rotor diameter. To translate this to grid sites, we use 
\begin{equation}
    L = \left\lfloor \frac{\text{Side length}}{\text{Resolution}} \right\rfloor + 1,
\end{equation}
where the side length and resolution are given in meters.

\subsection{Scaling-test toy model}\label{sec:TM}
\begin{figure}[H]
  \centering
  \begin{subfigure}[t]{0.48\textwidth}
    \centering
    \includegraphics[width=\linewidth]{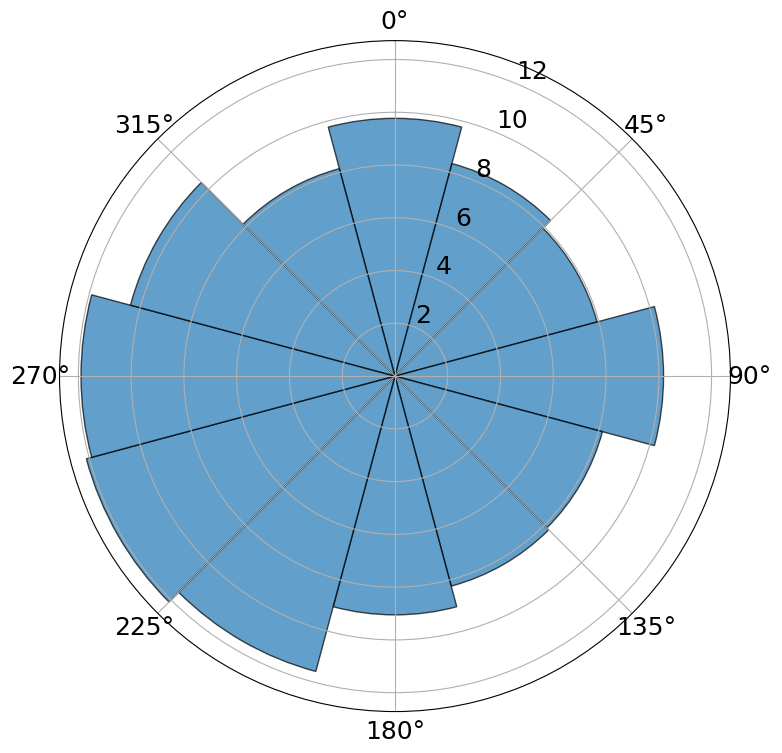}
    \caption{Rose diagram of North Sea data windspeeds.}
    \label{fig:theirRoseSpeed}
  \end{subfigure}%
  \hfill
  \begin{subfigure}[t]{0.48\textwidth}
    \centering
    \includegraphics[width=\linewidth]{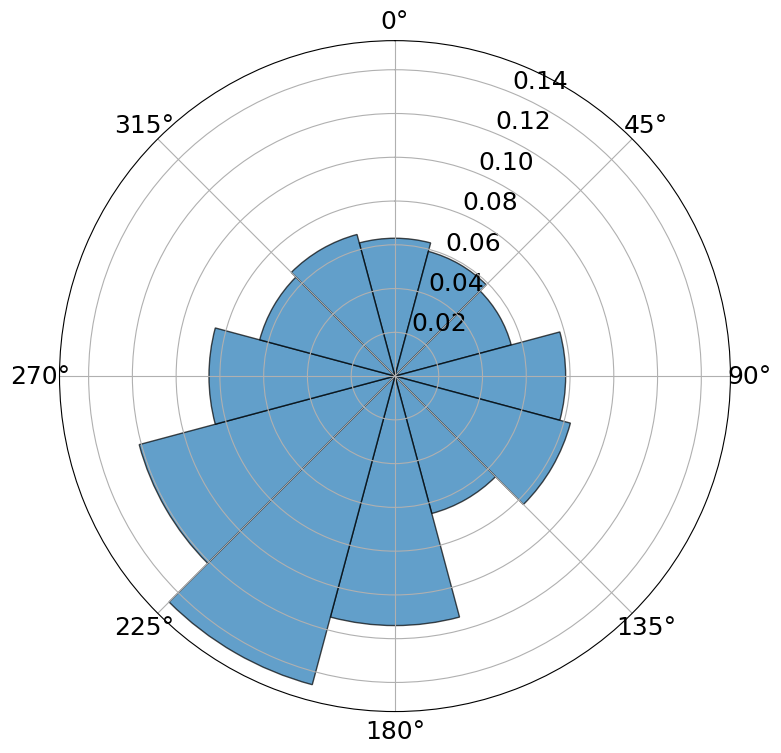}
    \caption{Rose diagram of North Sea data probabilities.}
    \label{fig:theirRoseProb}
  \end{subfigure}
  \caption{Rose diagrams showing windspeed and probability distributions for North Sea data.}
  \label{fig:theirRoseCombined}
\end{figure}

The model constructed by Rodrigues et al. makes use of a wind regime based on data collected in the North Sea \cite{NSWD} (shown in \tref{tab:literature_windregime}) for their offshore windfarm simulations. This data can also be represented as the two rose diagrams, \fref{fig:theirRoseSpeed} and \fref{fig:theirRoseProb}.
\begin{table}[h!]
    \centering
    \begin{tabular}{lrrrrrr}
        \toprule
        $\alpha_d$ & $v_d$ & $p_d$ \\
        \midrule
        $0$ & 9.77 & 0.063 \\
        $30$ & 8.34 & 0.059 \\
        $60$ & 7.93 & 0.055 \\
        $90$ & 10.18 & 0.078 \\
        $120$ & 8.14 & 0.083 \\
        $150$ & 8.24 & 0.065 \\
        $180$ & 9.05 & 0.114 \\
        $210$ & 11.59 & 0.146 \\
        $240$ & 12.11 & 0.121 \\
        $270$ & 11.90 & 0.085 \\
        $300$ & 10.38 & 0.064 \\
        $330$ & 8.14 & 0.067 \\
        \bottomrule
    \end{tabular}
    \caption{North Sea data wind regime. $\alpha_d$ is the angle, $v_d$ is the free windspeed and $p_d$ is the probability of a given arrangement $d$.}
    \label{tab:literature_windregime}
\end{table}

\Tref{tab:literature_scaling} shows the different windfarms that were tested by Rodrigues et al., and how these are translated into our model parameters.

\begin{figure}[H]
\centering
\includegraphics[width=0.6\linewidth]{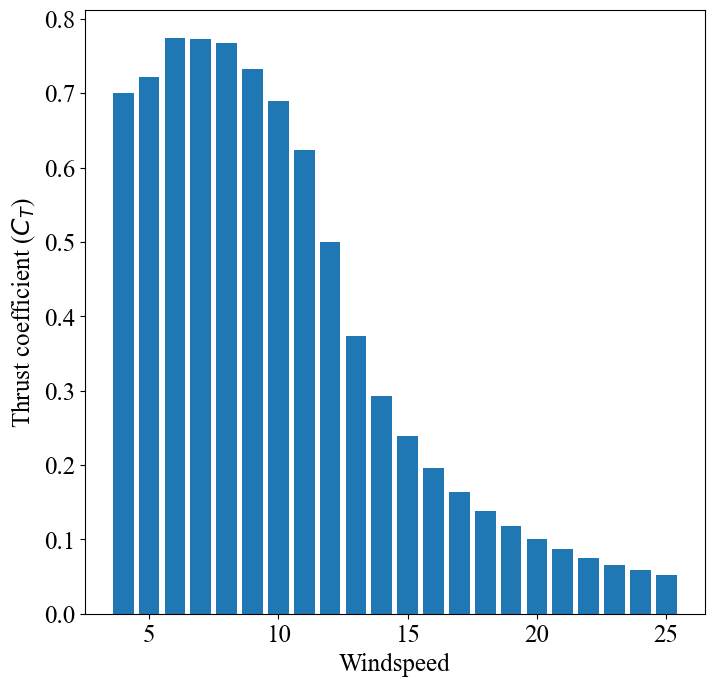}
\caption{Thrust coefficient, $C_T$, at different free windspeeds for the turbines used in Reference \cite{rod2016}.}
\label{fig:literature_power2}
\end{figure}

\begin{table*}[h!]
    \centering
    \begin{tabular}{lrrrrr}
        \toprule
        Farm Label & $W_A$ (km$^2$) & $W_L$ (m) & $M$ & Resolutions & $L$s\\ 
        \midrule
        $A$ & 15.49 & 3940 & 16 & $8D$, $4D$, $2D$ & 4, 7, 13\\
        $B$ & 61.97 & 7872 & 49 & $8D$, $4D$, $2D$ & 7, 13, 25\\
        $C$ & 139.43 & 11808 & 100 & $8D$, $4D$, $2D$ & 10, 19, 37\\
        $D$ & 247.89 & 15745 & 169 & $8D$, $4D$, $2D$ & 13, 25, 49\\
        \bottomrule
    \end{tabular}
    \caption{Different windfarms tested by Rodrigues et al. and the corresponding model parameters.}
    \label{tab:literature_scaling}
\end{table*}
\FloatBarrier
The choice of turbine influences both the wake dynamics and the power output of a windfarm. The authors of reference \cite{rod2016} opted for a turbine with a diameter of $D=2r_t = 164$m and a hub height of $h=107$m. To calculate the power output of the turbines at different windspeeds, we can refer to \tref{tab:literature_power}, also shown in \fref{fig:literature_power2}. 
\begin{table}[H]
\centering
\begin{tabular}{lrr}
\toprule
Windspeed & $C_T$ \\
\midrule
4 & 0.7000000000 \\
5 & 0.722386304 \\
6 & 0.773588333 \\
7 & 0.773285946 \\
8 & 0.767899317 \\
9 & 0.732727569 \\
10 & 0.688896343 \\
11 & 0.623028669 \\
12 & 0.500046699 \\
13 & 0.373661747 \\
14 & 0.293230676 \\
15 & 0.238407400 \\
16 & 0.196441644 \\
17 & 0.163774674 \\
18 & 0.137967245 \\
19 & 0.117309371 \\
20 & 0.100578122 \\
21 & 0.086883163 \\
22 & 0.075565832 \\
23 & 0.066131748 \\
24 & 0.058204932 \\
25 & 0.051495998 \\
\bottomrule
\end{tabular}
\caption{Thrust coefficient, $C_T$, at different free windspeeds for the turbines used in Reference \cite{rod2016}.}
\label{tab:literature_power}
\end{table}
\FloatBarrier
We set the wake expansion factor via
\begin{equation}
    a = \frac{A}{\log(\text{Hub height}/z_0)} = \frac{0.5}{\log(107/0.0005)} = 0.094,
\end{equation}
where $A$ is a constant and $z_0$ is the surface roughness. This value is in terms of meters.

Although Rodrigues et al. use the Katic-Jensen wake model to calculate the wake interactions (details of which can be found in appendix \ref{appen:KJWM}),  we instead make use of the same wake model as defined in \sref{sec:WFLOQUBO} due to formulation restrictions. For a fair baseline test, we replicate the original authors' algorithm by omitting the explicit minimum spacing constraint ($E$) and the set of prohibited locations ($\vec{P}$).

Using these features, we construct models suitable for scaling tests. As indicated in \tref{tab:literature_scaling} \cite{rod2016}, many benchmark problems are large-scale, with reported run times up to approximately 110 hours (see \cite[Fig. 15, p. 604]{rod2016}). These times would be longer for quantum simulation, with significantly higher memory overhead. To limit computational and energy usage, we will use only Windfarms $A$ and $B$ for comparison. Using the SQOE mapping, this will not affect the qubit count, as this is something we can freely choose. However, when using the PCE, the system with $L=13$ would require $N=169$ binary variables, and in turn, $12$-$57$ qubits, depending on choice of $k$. Due to these technical limitations, we test up to a maximum grid size of $9\times 9$.

Explicitly, we will test the systems shown in \tref{tab:my_scaling}.

\begin{table}[H]
    \centering
    \begin{tabular}{lrrrrr}
        \toprule
        Farm Label & $W_A$ (km$^2$) & Side length (m) & $M$ & $L$s\\ 
        \midrule
        $A$ & 15.49 & 3940 & 16  & 4, 7, 9\\
        $B$ & 61.97 & 7872 & 49 & 7, 9\\
        \bottomrule
    \end{tabular}
    \caption{Different windfarms that we will test and their corresponding model parameters.}
    \label{tab:my_scaling}
\end{table}

\subsection{Real model}
To evaluate all features of our models, we test it on the real-world Alltwalis windfarm \cite{statkraft_alltwalis} in Wales. We conduct this analysis at multiple scales to assess the model's performance and scalability under realistic conditions. We construct the model for $L = 7$, $8$, $9$. This requires several building blocks at each scale - namely: choosing $M$, calculating $E$, and constructing $\vec{P}$. We also need to construct a full wind regime, $D$, from available wind pattern data.

\begin{figure}[H]
    \centering
    \includegraphics[width=0.7\linewidth]{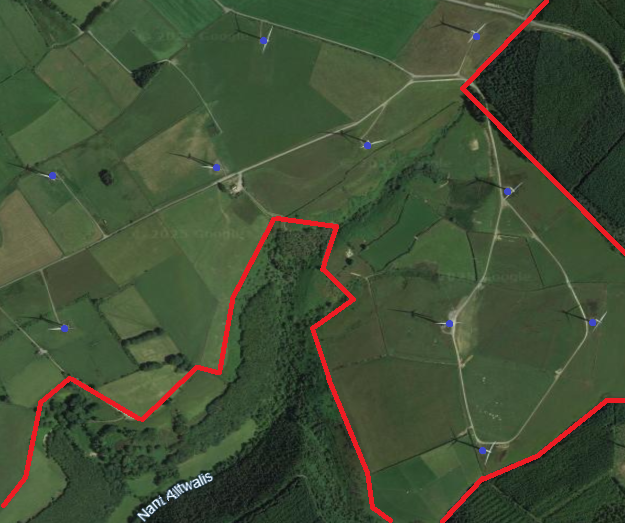}
    \caption{Highlighted image of Alltwalis farm. Areas where turbines cannot be placed are highlighted in red, and the actual turbine locations are marked in blue.}
    \label{fig:highlighted}
\end{figure}

\begin{figure}[H]
  \centering
  \begin{subfigure}[t]{0.48\textwidth}
    \centering
    \includegraphics[width=\linewidth]{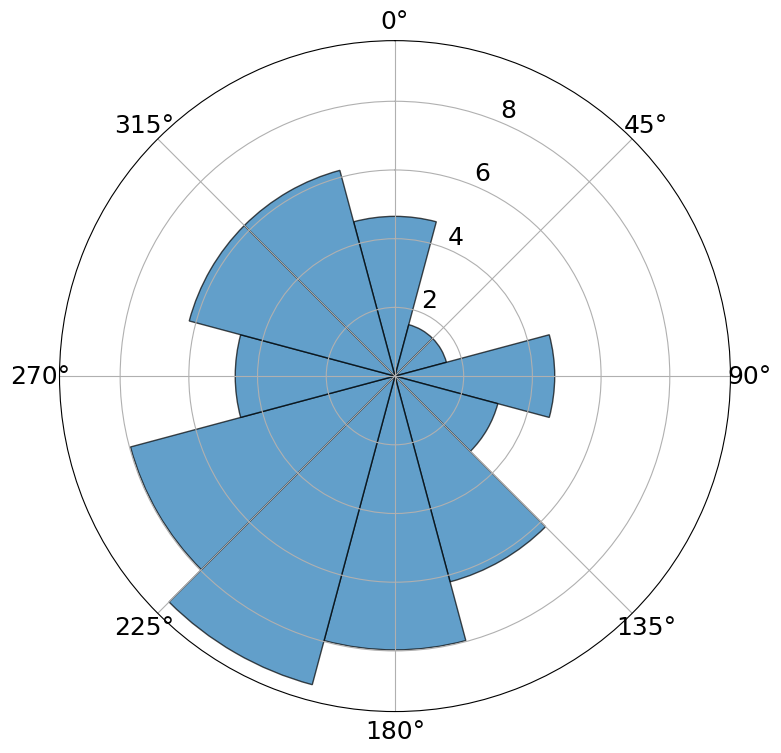}
    \caption{Rose diagram of Alltwalis windfarm data windspeeds.}
    \label{fig:alltRoseSpeed}
  \end{subfigure}%
  \hfill
  \begin{subfigure}[t]{0.48\textwidth}
    \centering
    \includegraphics[width=\linewidth]{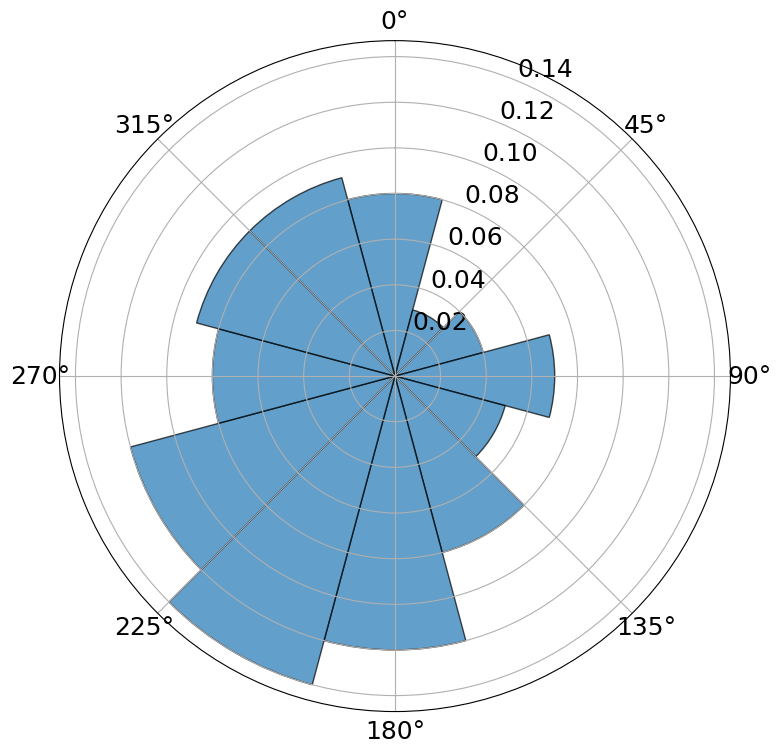}
    \caption{Rose diagram of Alltwalis windfarm data probabilities.}
    \label{fig:alltRoseProb}
  \end{subfigure}
  \caption{Rose diagrams showing windspeed and probability distributions for Alltwalis windfarm.}
\end{figure}

\Fref{fig:highlighted} is a Google Maps image of the farm, with areas unsuitable for turbine placement (due to terrain) highlighted in red and the actual turbine locations marked in blue. The field has an area of approximately $W_A=2.5\text{km}^2$, corresponding to a side length of $W_L=1581.13$\,m. The wind regime is shown in \tref{tab:allt_windregime}, and in rose diagram form in \fref{fig:alltRoseSpeed} and \fref{fig:alltRoseProb}. Data for the wind regime was taken from reference \cite{GlobalWindAtlas2025}. Appendix \ref{appen:P} shows our estimated choices for $\vec{P}$ at each scale based on \fref{fig:highlighted}. 

\begin{table}[H]
    \centering
    \begin{tabular}{lrr}
        \toprule
        $\alpha_d$ & $v_d$ & $p_d$ \\
        \midrule
        $0$ & 4.65 & 0.08 \\
        $30$ & 1.55 & 0.03 \\
        $60$ & 1.55 & 0.04 \\
        $90$ & 4.65 & 0.07 \\
        $120$ & 3.10 & 0.05 \\
        $150$ & 6.20 & 0.08 \\
        $180$ & 7.97 & 0.12 \\
        $210$ & 9.30 & 0.14 \\
        $240$ & 7.97 & 0.12 \\
        $270$ & 4.65 & 0.08 \\
        $300$ & 6.20 & 0.09 \\
        $330$ & 6.20 & 0.09 \\
        \bottomrule
    \end{tabular}
    \caption{Alltwalis wind regime data from reference \cite{GlobalWindAtlas2025}. $\alpha_d$ is the angle, $v_d$ is the free windspeed, and $p_d$ is the probability of each wind arrangement $d$.}
    \label{tab:allt_windregime}
\end{table}

\begin{figure}[H]
    \centering
    \includegraphics[width=0.6\linewidth]{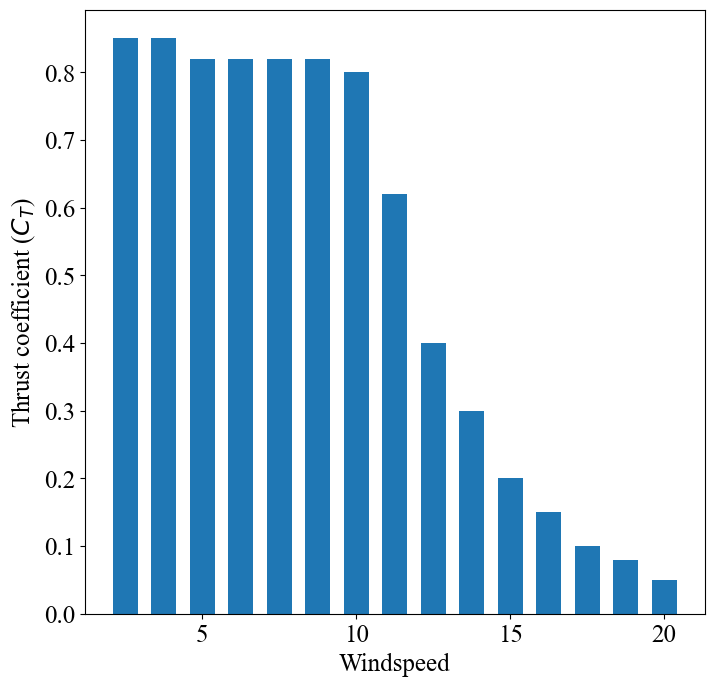}
    \caption{Thrust coefficient, $C_T$, at different free windspeeds for the turbines at Alltwalis windfarm.}
    \label{fig:my_power}
\end{figure}

The physical features of the turbines at Alltwalis allow us to build an accurate wake model - this windfarm has ten Siemens SWT-2.3-93 turbines \cite{thewindpower_alltwalis_2025}. These turbines have a rotor diameter of 93m, and a hub height of approximately 90m \cite{siemans}. Thrust data, taken from the figure 2(a) from reference \cite{wes-5-1551-2020}, is shown in \tref{tab:my_power} and \fref{fig:my_power}.
\begin{table}[H]
    \centering
    \begin{tabular}{lr}
        \toprule
        Windspeed & $C_T$ \\
        \midrule
        2.5 & 0.85 \\
        3.75 & 0.85 \\
        5.0 & 0.82 \\
        6.25 & 0.82 \\
        7.5 & 0.82 \\
        8.75 & 0.82 \\
        10.0 & 0.8 \\
        11.25 & 0.62 \\
        12.5 & 0.4 \\
        13.75 & 0.3 \\
        15.0 & 0.2 \\
        16.25 & 0.15 \\
        17.5 & 0.1 \\
        18.75 & 0.08 \\
        20.0 & 0.05 \\
        \bottomrule
    \end{tabular}
    \caption{Thrust coefficient, $C_T$, at different free windspeeds for the turbines at Alltwalis windfarm.}
    \label{tab:my_power}
\end{table}

The wake expansion factor of this onshore windfarm can be calculated as
\begin{equation}
    a = \frac{A}{\log(\text{Hub height}/z_0)} = \frac{0.5}{\log(90/0.05)} = 0.154,
\end{equation}
where $A$ can be set to $0.5$ as before, and $z_0$, the surface roughness, can be set to 0.05 \cite{burton2011wind}. Alltwalis windfarm is among mainly farmland with some trees, and so this value is appropriate.

Referring again to the Wind Energy Handbook \cite{burton2011wind}, an appropriate value for the minimum spacing $E$ is approximately $5D = 10r_t = 465$ m.

\Tref{tab:summary} contains all relevant model parameters that we test in this study.

\begin{table*}[h!]
    \centering
    \hspace*{-0.8cm}
    \begin{tabular}{lrrrrr}
        \toprule
        Model & $r_t$ (m) & $h$ (m) & Wake equation (m) & $W_A$ (km$^2$)\\
        \midrule
        Windfarm $A$ &  82 & 107 & $r_w=82 + 0.094\delta$ & 15.49\\
        Windfarm $B$ &  82 & 107 & $r_w=82 + 0.094\delta$ & 61.97\\
        Alltwalis & 46.5 & 90 & $r_w=46.5 + 0.154\delta$ & 1.5\\
        \bottomrule
    \end{tabular}
    
    \vspace{0.5cm}
    
    \begin{tabular}{lrrrrr}
        \toprule
        Model & Wind regime & $L$ & $M$ & $E$ (m) & $\vec{P}$ \\
        \midrule
        Windfarm $A$ & \Tref{tab:literature_windregime} & 4, 7, 9 & 16 & - & - \\
        Windfarm $B$ & \Tref{tab:literature_windregime} & 7, 9 & 49 & - & - \\
        Alltwalis & \Tref{tab:allt_windregime} & 7, 8, 9 & 10 & 465 & \ref{appen:P}\\
        \bottomrule
    \end{tabular}
    \caption{Table of parameters for our test models. Windfarm $A$ and $B$ are from reference \cite{rod2016}, details on the Alltwalis farm can be found in reference \cite{thewindpower_alltwalis_2025}. Here, $r_t$ is the turbine radius, $h$ is the hub height, and $W_A$ is the area of the windfarm. \Tref{tab:literature_windregime} contains the North sea wind data and \tref{tab:allt_windregime} the Alltwalis data.}
    \label{tab:summary}
\end{table*}

\section{Choices for $\vec{P}$ for Alltwalis Windfarm}\label{appen:P}
\begin{figure}[H]
  \centering
  \begin{subfigure}[t]{0.32\textwidth}
    \centering
    \includegraphics[width=\linewidth]{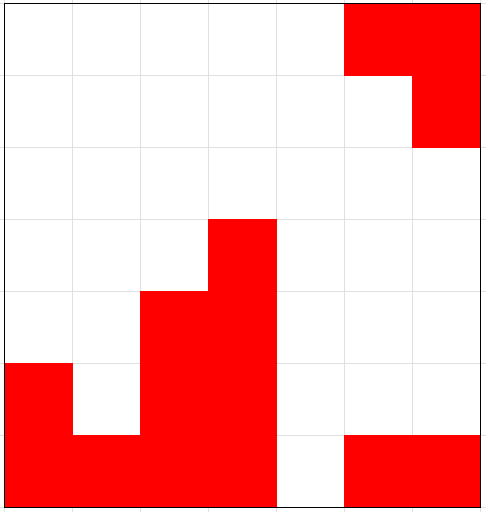}
    \caption{$L = 7$}
    \label{fig:p7}
  \end{subfigure}%
  \hfill
  \begin{subfigure}[t]{0.32\textwidth}
    \centering
    \includegraphics[width=\linewidth]{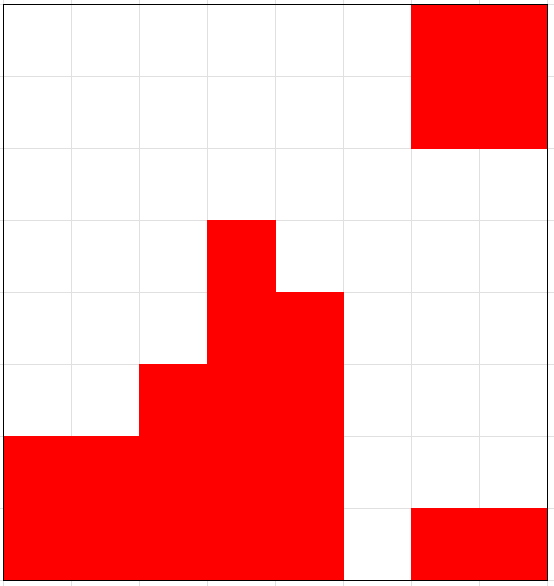}
    \caption{$L = 8$}
    \label{fig:p8}
  \end{subfigure}%
  \hfill
  \begin{subfigure}[t]{0.32\textwidth}
    \centering
    \includegraphics[width=\linewidth]{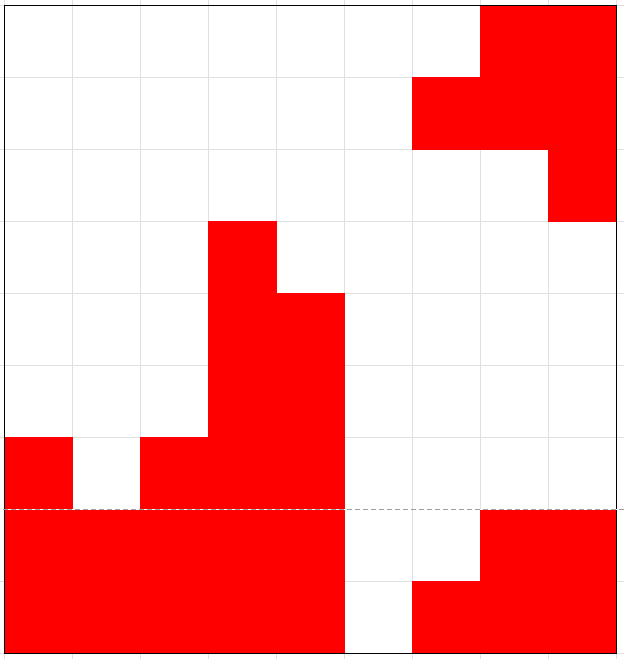}
    \caption{$L = 9$}
    \label{fig:p9}
  \end{subfigure}
  
  \caption{$\vec{P}$ at different scales.}
  \label{fig:pscales}
\end{figure}

The unusable areas of the Alltwalis wind farm, approximated using grid sizes of L = 7, 8, and 9, are shown in \fref{fig:p7}, \fref{fig:p8}, and \fref{fig:p9}, respectively.

\section{Constraint Weight Justification}\label{appen:Lambda}
Windfarm $A$, from \tref{tab:summary} in appendix \ref{appen:TM}, is used to construct the matrix for this analysis. The value of $\lambda$ is required to be large enough to meet the problem constraints, while prioritizing high-quality placement according to the wind regime. We analyze two key features: the cost gap (cost difference between the optimal solution(s) and the next-best solution(s)), and the number of turbines in the optimal layout.

\begin{figure}[H]
    \centering

    \begin{subfigure}[t]{0.45\linewidth}
        \centering
        \includegraphics[width=\linewidth]{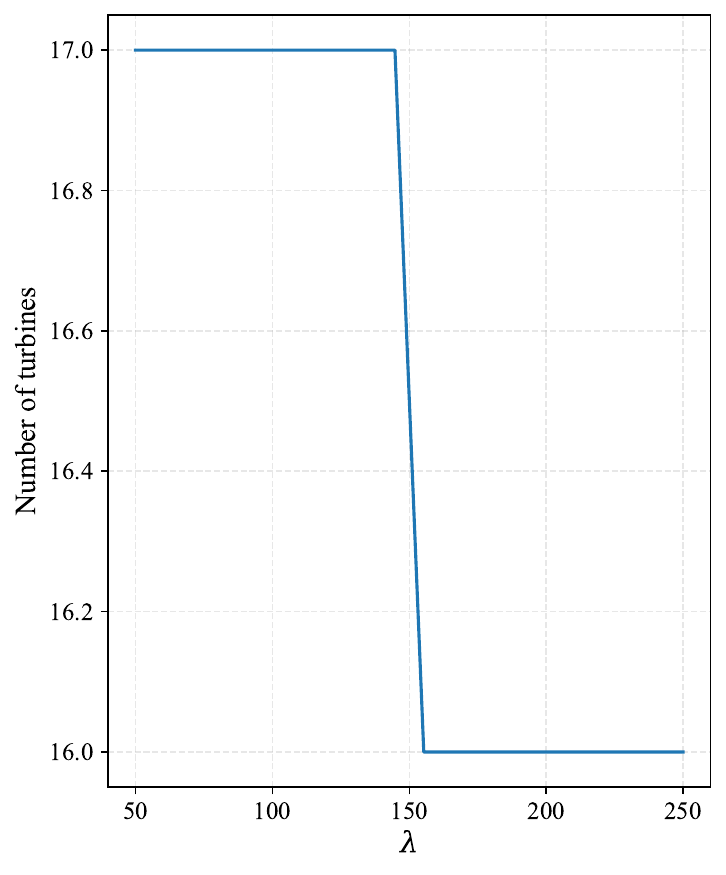}
        \caption{The number of turbines in the optimal solution for a range of constraint weight values. This transitions at $\lambda = 150$.}
        \label{fig:turbinesOptimals}
    \end{subfigure}%
    \hfill
    \begin{subfigure}[t]{0.45\linewidth}
        \centering
        \includegraphics[width=\linewidth]{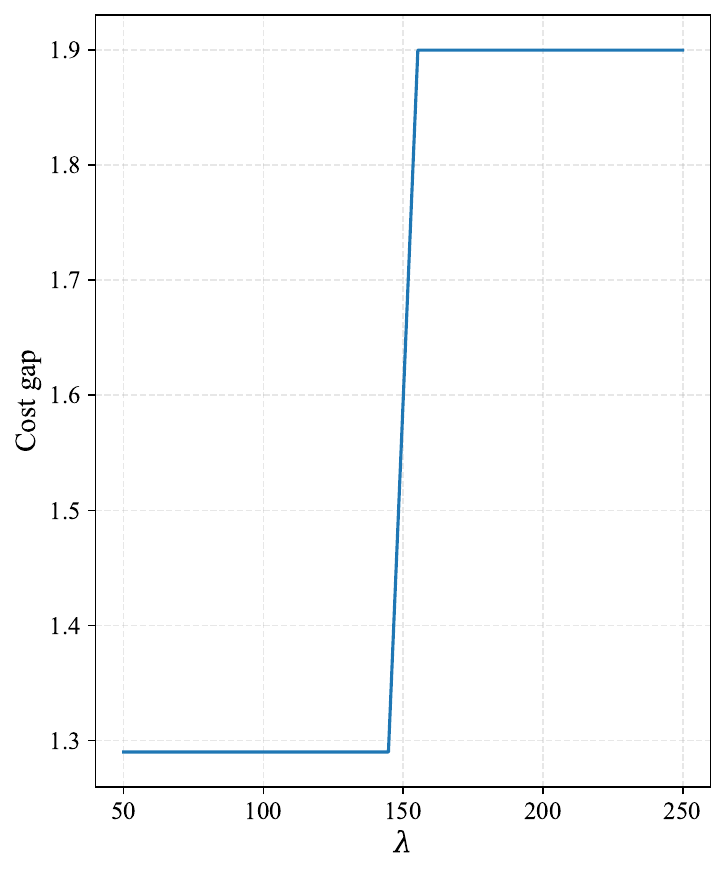}
        \caption{The cost gap between the optimal solution and the second best for a range of constraint weight values. This transitions at $\lambda = 150$.}
        \label{fig:gapOptimals}
    \end{subfigure}
    \caption{Scan over suitable choices for constraint weight $\lambda$. In both cases, we see that for $\lambda > 150$ the optimal solution changes to meet the constraints.}
\end{figure}

For any value larger than 150, the optimal solution meets the constraints, and the gap between the optimal and next best solution(s) steps to a constant value (\fref{fig:turbinesOptimals} and  \fref{fig:gapOptimals}); we opt for $\lambda > 190$. 

\section{Katic-Jensen Wake Model}\label{appen:KJWM}
The Katic-Jensen (K-J) wake model is a modification of the original Jensen wake model \cite{jensen1983note}, by Katic et al. \cite{3a81166868144671af170595fd17b8f6}. Similar to the JM, K-J is defined as
\begin{equation}
    u_{j} = v_d(1-\text{deficit)},
\end{equation}
To find the deficit, we now calculate a set of values $U_{kj}$ - the interference at the turbine $j$ due to the presence of a turbine at $k$. $U_{kj}$ can be calculated via
\begin{equation}\label{eq:Ukj}
    U_{kj} = \left(1-\sqrt{1-C_T}\right)\left(1+\frac{a \delta}{r_t}\right)^{-2} \frac{A_{kj}}{A_r}.
\end{equation}
From here
\begin{equation}
    \text{deficit} = \sqrt{\sum_{k=1}^nU_{kj}^2}
\end{equation}
here, $C_T$ is the thrust coefficient, $a$ is the wake expansion factor, $\delta$ is the Euclidean distance between the turbines, and $A_r = \pi r_t^2=\pi (D/2)^2$ is the turbine rotor area. The (effective) turbine rotor area $A_{kj}$ influenced by the turbine at $k$ is calculated as
\begin{equation}\label{eq:Akj}
\begin{aligned}
A_{kj} &= \frac{1}{2} \left( r_w^2 \left( 2\arccos \left( \frac{r_w^2 + \delta^2 - r_t^2}{2r_w \delta} \right) - \sin \left( 2\arccos \left( \frac{r_w^2 + \delta^2 - r_t^2}{2r_w \delta} \right) \right) \right) \right) \\
&\quad + \frac{1}{2} \left( r_t^2 \left( 2\arccos \left( \frac{r_t^2 + \delta^2 - r_w^2}{2r_t \delta} \right) - \sin \left( 2\arccos \left( \frac{r_t^2 + \delta^2 - r_w^2}{2r_t \delta} \right) \right) \right) \right),
\end{aligned}
\end{equation}
where $r_w = r_t + a\delta$ is the wake radius, and $r_t = D/2$ is the turbine radius.

\end{document}